\documentclass[journal]{IEEEtran}

\ifCLASSINFOpdf
\else

\fi

\usepackage{cite}
\usepackage{lscape}
\usepackage{textcomp}
\usepackage{multirow}
\usepackage[graphicx]{realboxes}
\usepackage{adjustbox}
\usepackage{tabularx}
\usepackage{amsmath}
\usepackage{amssymb}
\usepackage[normalem]{ulem}
\usepackage{adjustbox}
\usepackage{tablefootnote}
\usepackage{xcolor}
\usepackage{verbatim}
\usepackage{graphicx}
\usepackage{float}
\usepackage{textgreek}

\newcommand{\blk}{\color{black}}

\makeatletter 
\pretocmd\@bibitem{\color{black}\csname keycolor#1\endcsname}{}{\fail}
\newcommand\citecolor[1]{\@namedef{keycolor#1}{\color{blue}}}
\makeatother

\useunder{\uline}{\ul}{}

\begin{document}
\newpage{
\onecolumn

\begin{center}
\begin{Huge}
\textbf{Notice:} \\

This work has been submitted to the IEEE for possible publication. Copyright may be transferred without notice, after which this version may no longer be accessible.

\end{Huge}
\end{center}
}

\twocolumn
\title{Local-Global Temporal Fusion Network with an Attention Mechanism for Multiple and Multiclass Arrhythmia Classification} 
\author{Yun Kwan Kim, Minji Lee, Kunwook Jo, Hee Seok Song, and Seong-Whan~Lee, \IEEEmembership{Fellow, IEEE}

\thanks{\textit{(Corresponding authors: Hee Seok Song, Seong-Whan Lee.)}}

\thanks{Y. K. Kim is with the Department of Brain and Cognitive Engineering, Korea University, 145, Anam-ro, Seongbuk-gu, Seoul 02841, and the Technology Development, Seers Technology Co. Ltd., 1401, 8, Seongnam-daero 331 beon-gil, Bundang-gu, Seongnam-si, Gyeonggi-do 13558, Republic of Korea (e-mail: ykwin@korea.ac.kr).}
\thanks{M. Lee is with the Department of Biomedical Engineering, The Catholic University of Korea, 43, Jibong-ro, Bucheon-si, Gyeonggi-do 14662, Republic of Korea (e-mail:  minjilee@catholic.ac.kr).}
\thanks{K. Jo and H. S. Song are with the Technology Development, Seers Technology Co. Ltd., 1401, 8, Seongnam-daero 331beon-gil, Bundang-gu, Seongnam-si, Gyeonggi-do 13558, Republic of Korea (e-mail: tom.jo@seerstech.com; sam.song@seerstech.com).}
\thanks{S.-W. Lee is with the Department of Artificial Intelligence, Korea University, 145, Anam-ro, Seongbuk-gu, Seoul 02841, Republic of Korea (e-mail: sw.lee@korea.ac.kr).}
}

\markboth{}
{Kim \MakeLowercase{\textit{et al.}}: Local-Global Temporal Fusion Network with Attention Mechanism for Multiple and Multiclass Arrhythmia Classification}

\maketitle

\begin{abstract}
Clinical decision support systems (CDSSs) have been widely utilized to support the decisions made by cardiologists when detecting and classifying arrhythmia from electrocardiograms (ECGs). However, forming a CDSS for the arrhythmia classification task is challenging due to the varying lengths of arrhythmias. Although the onset time of arrhythmia varies, previously developed methods have not considered such conditions. Thus, we propose a framework that consists of (i) local temporal information extraction, (ii) global pattern extraction, and (iii) local-global information fusion with attention to perform arrhythmia detection and classification with a constrained input length. The 10-class and 4-class performances of our approach were assessed by detecting the onset and offset of arrhythmia as an episode and the duration of arrhythmia based on the MIT-BIH arrhythmia database (MITDB) and MIT-BIH atrial fibrillation database (AFDB), respectively. The results were statistically superior to those achieved by the comparison models. To check the generalization ability of the proposed method, an AFDB-trained model was tested on the MITDB, and superior performance was attained compared with that of a state-of-the-art model. The proposed method can capture local-global information and dynamics without incurring information losses. Therefore, arrhythmias can be recognized more accurately, and their occurrence times can be calculated; thus, the clinical field can create more accurate treatment plans by using the proposed method.
\end{abstract}


\section{Introduction}
\label{sec:introduction}

\IEEEPARstart{C}linical decision support systems (CDSSs) are medical information technology systems designed to provide clinical decision support using a knowledge base; such systems apply rules to patient data and employ machine learning to analyze clinical data \cite{xu2021comorbidity, jeong2022real}. CDSSs have been developed using electrocardiograms (ECG) to take advantage of various paradigms, such as Holter monitoring systems, real-time patient monitoring, and heart failure prediction \cite{sutton2020overview}. Holter monitoring systems, which require careful examination of long ECG records for analysis purposes, have developed an arrhythmia detection and classification algorithm in the area of CDSSs to enhance the grades and speeds of medical services \cite{sutton2020overview}. These computer-aided arrhythmia classification algorithms offer new opportunities for cardiologists to achieve enhanced diagnostic accuracy \cite{sidek2014ecg}.

An ECG indicates the heart’s electrical movement over electrodes placed on the human hide. ECGs are commonly applied in clinical settings for arrhythmia detection. Arrhythmia has some defining characteristics, such as the fact that it occurs and then terminates abruptly \cite{kamel2013paroxysmal}. In addition, arrhythmia can occur for a long time \cite{atlee2006complications}. Cardiologists observe the pathology of the cardiovascular system through ECG rhythm changes and provide important reference information for cardiac diagnosis  \cite{sannino2018deep}. Holter recordings of over 24 h are essential in the field of CDSSs for accurately detecting arrhythmias. The manual inspection and interpretation of ECGs are time-consuming and burdensome processes, as ECGs they are recorded over a long period of time. To address these challenges, several studies have developed CDSSs using automated algorithms to achieve enhanced arrhythmia detection and diagnostic accuracy \cite{kim2022automatic, huang2014new}. 

Many studies have utilized machine learning methods using preprocessing, feature extraction, and feature segmentation techniques for automated arrhythmia classification \cite{pourbabaee2018deep}. To develop robust Holter monitoring systems for arrhythmia classification, many studies have used deep learning approaches \cite{kim2022automatic, pourbabaee2018deep}, leading to noteworthy performance improvements. For instance, Acharya et al. \cite{acharya2017automated} proposed a convolutional neural network (CNN) including eleven layers to classify four classes and achieved an F1 score of 83.00\%. By using a long-short term memory (LSTM) model, Faust et al. \cite{faust2018automated} obtained an F1 score of 99.77\% when identifying and classifying two types of arrhythmia. Chen et al. \cite{chen2020automated} suggested a method combining a CNN with LSTM to identify six types of arrhythmia, and their approach achieved an F1 score of 90.82\%. Teplitzky et al. \cite{teplitzky2020deep} suggested a ResNet-based sequence-to-sequence architecture for 14-class, 11-class, 7-class, and 5-class arrhythmia classifications and obtained F1 score of 70\%, 80\%, 90\%, and 95\%, respectively.

\begin{figure}[t]
\begin{center}
\includegraphics[width = 1.0\columnwidth]{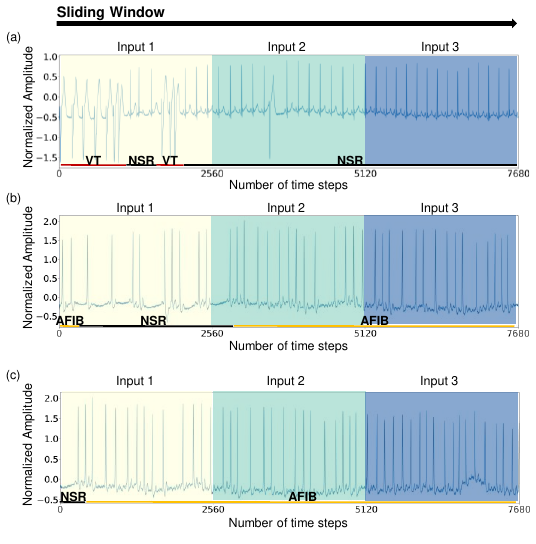}
\caption{Representative examples of arrhythmia characteristics from the MITDB. Each example shows the ECG signals of arrhythmia characteristics, including abrupt changes and persistent progression, when using the sliding window method. Each input window is 10 s long, and the color of each input is expressed differently. Next, at the bottom, the red horizontal bar indicates VT, the black horizontal bar represents NSR, and finally, the yellow horizontal bar shows AFIB. (a) VT and NSR. (b) AFIB and NSR. (c) NSR and AFIB.}
\end{center}
\end{figure}

Computer-aided CDSSs, including Holter monitoring and real-time arrhythmia monitoring systems in the real world, set a predefined input window and analyze the output in the corresponding input window. In addition, the computer-aided system in \cite{zhang2019online} used a sliding window that output its result by moving a predefined input window. These systems proceed with arrhythmia classification by assuming that there is one output for one input. However, a specific segmented input signal may include multiple events because of cardiac dynamics, as shown in Fig. 1.

These deep learning-based approaches have achieved good arrhythmia classification results, but Holter monitoring, real-time patient monitoring, and smart wearable devices using artificial intelligence (AI)-based applications still face challenges. Although previous studies have suggested methods for improving the effect of multiclass classification for one output, it is difficult to detect multiple arrhythmias in a predetermined window. Multiple arrhythmias exhibit a number of arrhythmia appearances as various ECG patterns (among those mentioned above) within a predetermined input window, as shown in Fig. 1. For example, in an input ECG signal of 10-s, ventricular tachycardia (VT) may appear within 0-4 s, NSR may appear within 4-6 s, VT may reappear within 6-8 s, and NSR may reappear within 8-10 s, as shown in (a) of Fig. 1. There are 4 events, including VT and NSR. This indicates that more than one arrhythmia event can occur within 10 s of input data. Moreover, “multiclass” means that each event is classified as one class among three or more classes \cite{aly2005survey}.

Many studies have focused on the detection of whether atrial fibrillation (AFIB) or non-AFIB exists in a fixed window \cite{parvaneh2019cardiac}. This task involves one input and multiple binary-class outputs. Although AFIB is the most common arrhythmia type in the adult population \cite{miyasaka2006secular}, various types of arrhythmia have recently been reported \cite{khurshid2018frequency}. It is important to detect and accurately classify various types of arrhythmia because these arrhythmias may transition to heart attacks, heart failure, and strokes \cite{walsh2007arrhythmias}. Therefore, long-term Holter monitoring and real-time monitoring within CDSS must detect and classify these arrhythmias as well as AFIB \cite{sutton2020overview}. However, detecting and extracting dynamic cardiac patterns and classifying arrhythmia in a detected cardiac pattern is challenging.

This study aimed to address these problems by suggesting a novel framework for arrhythmia detection and classification in a CDSS; this approach has the ability to improve its ability to capture temporal information dependencies via the fusion of local and global temporal patterns. Our proposed framework consists of an encoder-decoder architecture \cite{chaurasia2017linknet}, a temporal convolution network (TCN) \cite{lea2017temporal} with multiscale temporal information fusion (TIF), a dilated convolution layer \cite{yu2015multi}, and a multihead self-attention (MHA) mechanism \cite{vaswani2017attention}. As the baseline architecture, we used LinkNet \cite{chaurasia2017linknet}, which is a semantic segmentation network that utilizes encoder-decoder architectures, skip connections, and residual blocks. In addition, we suggested a mixture of the categorical cross-entropy (CCE) loss and the multiple weighted dice loss function to resolve class imbalances. The MIT-BIH arrhythmia database (MITDB) \cite{moody2001impact} and MIT-BIH AFIB database (AFDB) \cite{moody2001impact} were used. Also, the fusion of global flow information through multiscale TIF and local temporal interaction through the TCN block was proposed to overcome the limitation of not being able to simultaneously detect positional and contextual information. Next, we proposed a fusion strategy by combining the temporal MHA in the encoder with the upsampled features in the decoder. Features for emphasizing the interaction of temporal information through the MHA were added with the features restored through the decoder to enrich the details of much contextual information.

The following is a summary of the study’s main contributions.

$\bullet$ We proposed a framework to simultaneously detect and classify multiple arrhythmias in one input by combining a TCN \cite{lea2017temporal} with a multiscale TIF module and an MHA mechanism for extracting and fusing the temporal dynamics and positional information of arrhythmias.

$\bullet$ The fusion strategy was designed for the detection and classification of arrhythmia using two functions, where a feature map that uses the self-attention mechanism of TIF and long-term temporal relationship steps without the loss of information were coupled to the decoder. Consequently, the proposed framework is a significant enhancement over previously developed models in terms of duration and episode performance.

$\bullet$ Our proposed framework simultaneously determines and classifies the onset and offset of arrhythmia concerning local cardiac dynamics and global arrhythmia flows; thus, the proposed framework can inform cardiologists of the duration and episodes of each instance of arrhythmia.

The rest of this study is divided into the following sections. In Sections 2 and 3, we introduce the related studies and our model, respectively. The experiments and validation are presented in Section 4. The results and overall study are covered in Section 5. The research is concluded in Section 6.

\section{Related Work}
For the classification of arrhythmia, several machine learning studies have been developed using ECG data. Moreover, detection and classification methods have been used to find onsets and offsets within records.

Acharya et al. \cite{acharya2017automated} suggested a CNN consisting of an eleven-layer network to categorize arrhythmia. They used a database that combined the AFDB, the MITDB, and the Creighton University Ventricular Tachyarrhythmia Database to consider four classes: AFIB, atrial flutter (AFL), ventricular fibrillation (VF), and NSR. The overall accuracy, sensitivity, and specificity achieved in this study were 94.0\%, 99.0\%, and 81.0\%, respectively.

\begin{figure*}
\begin{center}
\includegraphics[width=0.75\textwidth]{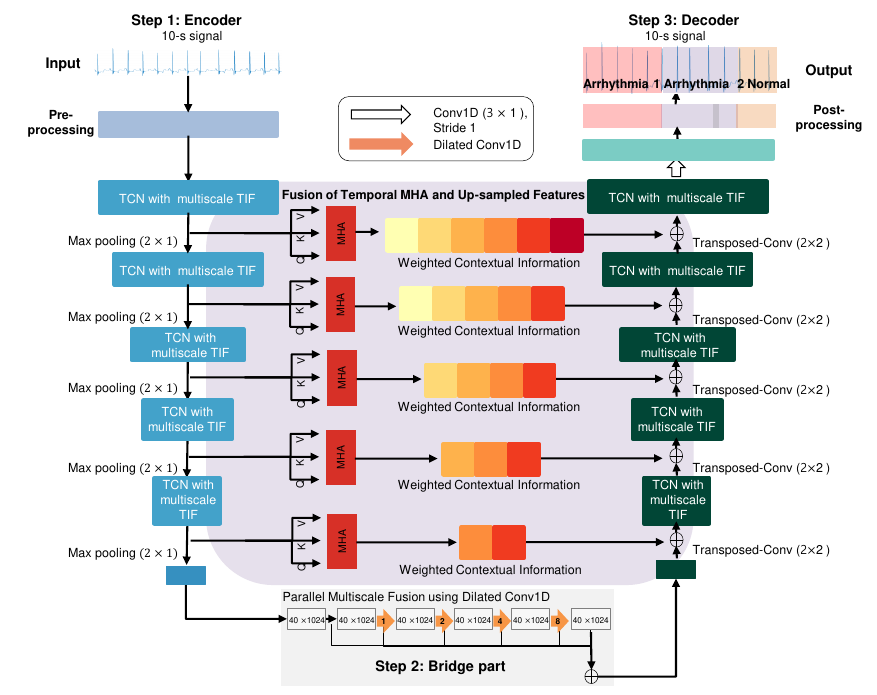}
\caption{\blk Overview of our arrhythmia detection and classification framework. Our suggested framework is composed of three steps. Step 1 encodes information by compressing features in the input data. The input in the encoder is made of a 10-s signal. The horizontal axis indicates the amplitude, while the vertical axis represents time. The normalized data are then inserted into five TCN blocks to extract features. A TCN block consists of two parts: a TCN and a multiscale TIF block. These features are then summed. The dilated Conv1D in each TCN block is used to handle long sequences. The dilation of $D$ is set to 1, 2, 3, 4, and 5. The output of each TCN block is passed to the corresponding decoder after applying the MHA. The weighted contextual information passed to the decoder fuses the upsampled information in the TCN with the multiscale TIF information of the decoder. A set containing $Q$, $K$, and $V$ is determined using the MHA technique based on scaled dot-product attention. Next, the information derived from the last residual block is fed to the parallel multiscale block using dilated Conv1D and then applied in Step 2. The bridge part includes a dilated Conv1D of the shortcut connections to form a parallel multiscale fusion structure for more comprehensively understanding the meaning of the arrhythmia information. Step 3 is composed of five TCN blocks with multiscale TIF and a transposed convolution layer. These layers improve the representation ability of the network. Temporal convolutional network: TCN. Temporal information fusion: TIF. One-dimensional dilated causal convolution: dilated Conv1D. Multihead self-attention: MHA. Queries: $Q$. Keys: $K$. Values: $V$.} \blk
\end{center}
\end{figure*}

Since LSTM is efficient at learning the sequential characteristics of ECG signals, it has been applied to classify arrhythmia. Faust et al. \cite{faust2018automated} suggested an LSTM approach for the detection and classification of AFIB and NSR. The only characteristic used in the CDSS of this model for AFIB was the time interval between two consecutive R peaks (the RR interval). On the AFDB, they achieved an F1 score of 99.8\%.

For ECG categorization, the combination of a CNN and LSTM is beneficial \cite{khalifa2019character}. To handle the ECG segments and accompanying RR intervals in the MITDB, Chen et al. \cite{chen2020automated} presented a network combination with a CNN and LSTM that used a multi-input structure. This research identified four categories of ECG data: sinus bradycardia (SBR), ventricular bigeminy, AFIB, and AFL. They obtained an overall accuracy of 99.3\%. He et al. \cite{he2019automatic} suggested a method with two modules comprising a residual convolutional network and bidirectional LSTM. Nine types of arrhythmias were categorized using the suggested model. The overall F1 scores achieved in this study were 91.4\%, 87.9\%, 80.1\%, and 74.2\%, respectively.

Simultaneous segmentation and classification tasks using ECG signals are primarily focused on beat classification. Beat classification approaches have been proposed as follows. Petryshak et al. \cite{petryshak2021robust} proposed an architecture that consists of segmentation and classification parts using the MITDB. The segmentation architecture localizes the R peaks of normal and anomalous beats using a fully convolutional encoder-decoder architecture. The classification architecture uses the CardioIncNet architecture function and then classifies PVC. The study obtained overall F1 score of 99.51\% and 96.40\% in R peak detection and PVC classification scenarios, respectively.

A sequence-to-sequence framework is a method for detecting onsets and offsets and performing classification. This type of method has produced striking results in machine translation systems \cite{johnson2017google}. Teplitzky et al. \cite{teplitzky2020deep} suggested a ResNet-based sequence-to-sequence architecture for 14-class, 11-class, 7-class, and 5-class arrhythmia classification. This architecture reported performance that was assessed once every second. They obtained overall F1 scores that surpassed 70\% for 14-class arrhythmia, 80\% for 11-class arrhythmia, 90\% for 7-class arrhythmia, and 95\% for 5-class arrhythmia. Pokaprakarn et al. \cite{pokaprakarn2021sequence} proposed another sequence-to-sequence architecture involving combinations of CNNs and LSTM to classify 5 cardiac rhythms. The input of the architecture was 5 s in length, which resulted in an output consisting of a sequence of cardiac rhythm labels. This approach achieved an overall F1 score of 89.00\%. 

In addition, TCNs have also been successful in learning sequential ECG features, thus addressing the problems of LSTM \cite{bai2018empirical}. Ma et al. \cite{ma2021ecg} suggested a TCN method for categorizing AFIB, AFL, and NSR using the AFDB. This study obtained an overall accuracy of 98.7\%. Infolfsson et al. \cite{ingolfsson2021ecg} suggested a TCN with three blocks to classify five beats using the ECG5000 dataset. They demonstrated accuracy and balanced accuracy rates of 94.2\% and 89.0\%, respectively. 

In conclusion, although some studies have proposed various frameworks for arrhythmia classification, the existence of multiple arrhythmias in a given ECG input remains problematic. For clinical applications in the field, a framework that can detect and classify arrhythmias with various lengths should be proposed.

\blk
\section{Proposed Method}
\subsection{Overview}
We suggest a detection and classification network inspired by LinkNet \cite{chaurasia2017linknet}. 
As shown in Fig. 2, the proposed framework comprises three parts: an encoder, a bridge part, and a decoder. Step 1 encodes information by compressing the features in the input data. The input in the encoder is made of a 10-s signal. The normalized data acquired after applying preprocessing and normalization are inserted into the encoder to extract features. The encoder contains five TCN blocks \cite{bai2018empirical} with a multiscale TIF module inspired by Ibtehaz et al. \cite{ibtehaz2020multiresunet} and Peng et al. \cite{peng2017large}. The TCN plays a role in capturing local long-term sequential patterns. The multiscale TIF module is configured to globally extract the ECG signal pattern. These features are then summed. The output of each TCN block is passed to the corresponding decoder after applying the MHA. The weighted contextual information passed to the decoder fuses the upsampled information in the TCN with the multiscale TIF information of the decoder. Next, the bridge part includes a dilated convolution of the shortcut connections for more comprehensively understanding arrhythmia information. Similar to the encoder, Step 3 is composed of five TCN blocks with multiscale TIF and a transposed convolution layer. These layers improve the representation ability of the network. Next, we compute the compound loss function using the weighted Dice coefficient and the CCE loss after completing these steps. Last, postprocessing corrects the incorrectly output samples (less than 256 samples) in the arrhythmia 2 period, although the proposed method outputs different arrhythmia 1, arrhythmia 2, and normal rhythms, as shown in Fig. 2. Table \uppercase\expandafter{\romannumeral1} lists the number of filters, kernel size, stride, activation function, and output size in each layer.

\begin{table*}
\caption{Layer and Hyperparameter Information of the Proposed Model}
\renewcommand{\arraystretch}{1.2}
\resizebox{\textwidth}{!}{
\begin{tabular}{cccccccc}
\hline
\textbf{Type}                                                   & \textbf{No.} & \textbf{Layer}      & \textbf{No. of filters} & \textbf{Size of kernel} & \textbf{Stride} & \textbf{Activation function} & \textbf{Output shape} \\ \hline
\multirow{12}{*}{\textbf{Encoder}}                              & 1                             & Input                                & -                                           & -                                     & -                                & -                                             & 2560 $\times$ 1                        \\ \cline{2-8} 
                                                                & 2                             & TCN block with multiscale TIF module & 64                                          & Kernel: 3, Dilation rate: 1                      & 1                                & ReLU                                          & 2560 $\times$ 64                       \\ \cline{2-8} 
                                                                & 3                             & Max pooling                          & -                                           & -                                     & 2                                & -                                             & 1280 $\times$ 64                       \\ \cline{2-8} 
                                                                & 4                             & TCN block with multiscale TIF module & 128                                         & Kernel: 3, Dilation rate: 2                      & 1                                & ReLU                                          & 1280 $\times$ 128                      \\ \cline{2-8} 
                                                                & 5                             & Max pooling                          & -                                           & -                                     & 2                                & -                                             & 640 $\times$ 128                       \\ \cline{2-8} 
                                                                & 6                             & TCN block with multiscale TIF module & 256                                         & Kernel: 3, Dilation rate: 4                      & 1                                & ReLU                                          & 640 $\times$ 256                       \\ \cline{2-8} 
                                                                & 7                             & Max pooling                          & -                                           & -                                     & 2                                & -                                             & 320 $\times$ 256                       \\ \cline{2-8} 
                                                                & 8                             & TCN block with multiscale TIF module & 512                                         & Kernel: 3, Dilation rate: 8                      & 1                                & ReLU                                          & 320 $\times$ 512                       \\ \cline{2-8} 
                                                                & 9                             & Max pooling                          & -                                           & -                                     & 2                                & -                                             & 160 $\times$ 512                       \\ \cline{2-8} 
                                                                & 10                            & TCN block with multiscale TIF module & 1024                                        & Kernel: 3, Dilation rate: 16                     & 1                                & ReLU                                         & 160 $\times$ 1024                      \\ \cline{2-8} 
                                                                & 11                            & Max pooling                          & -                                           & -                                     & 2                                & -                                             & 80 $\times$ 1024                       \\ \cline{2-8} 
                                                                & 12                            & Max pooling                          & -                                           & -                                     & 2                                & -                                             & 40 $\times$ 1024                       \\ \hline
\textbf{\begin{tabular}[c]{@{}c@{}}Bridge \\ Part\end{tabular}} & 13                            & Bridge Block                         &      1024    & \begin{tabular}[c]{@{}c@{}} Path 1: [Kernel: 2, Dilation rate: 1, 2, 4, 8] \\ Path 2: [Kernel: 2, Dilation rate: 1, 2, 4] \\ Path 3: [Kernel: 2, Dilation rate: 1, 2] \\ Path 4: [Kernel: 2, Dilation rate: 1] \\ Addition (Path 1, Path 2, Path 3, Path 4, Layer 12) \end{tabular}   &    1  & - & 40 $\times$ 1024   \\ \hline
\multirow{17}{*}{\textbf{Decoder}}                              & 14                            & Transposed Conv                      & 1024                                        & 2                                     & 2                                & ReLU                                          & 80 $\times$ 1024                       \\ \cline{2-8} 
                                                                & 14                            & \multicolumn{5}{c}{Addition (Layer 14, MHA with layer 11)}                                                                                                                                                    & 80 $\times$ 1024                       \\ \cline{2-8} 
                                                                & 15                            & TCN block with multiscale TIF module & 1024                                        & Kernel: 3, Dilation rate: 1                      & 1                                & ReLU                                          & 80 $\times$ 1024                       \\ \cline{2-8} 
                                                                & 16                            & Transposed Conv                      & 512                                         & 2                                     & 2                                & ReLU                                          & 160 $\times$ 512                       \\ \cline{2-8} 
                                                                & 17                            & \multicolumn{5}{c}{Addition (Layer 16, MHA with layer 9)}                                                                                                                                                     & 160 $\times$ 512                       \\ \cline{2-8} 
                                                                & 18                            & TCN block with multiscale TIF module & 512                                         & Kernel: 3, Dilation rate: 2                      & 1                                & ReLU                                          & 160 $\times$ 512                       \\ \cline{2-8} 
                                                                & 19                            & Transposed Conv                      &    256                                         & 2                                     & 2                                & ReLU                                          & 320 $\times$ 256                       \\ \cline{2-8} 
                                                                & 20                            & \multicolumn{5}{c}{Addition (Layer 19, MHA with layer 7)}                                                                                                                                                     & 320 $\times$ 256                       \\ \cline{2-8} 
                                                                & 21                            & TCN block with multiscale TIF module & 256                                         & Kernel: 3, Dilation rate: 4                      & 1                                & ReLU                                          & 320 $\times$ 256                       \\ \cline{2-8} 
                                                                & 22                            & Transposed Conv                      & 128                                         & 2                                     & 2                                & ReLU                                          & 640 $\times$ 128                       \\ \cline{2-8} 
                                                                & 23                            & \multicolumn{6}{c}{Addition (Layer 22, MHA with layer 5)}                                                                                                                                                                                              \\ \cline{2-8} 
                                                                & 24                            & TCN block with multiscale TIF module & 128                                         & Kernel: 3, Dilation rate: 8                      & 1                                & ReLU                                          & 640 $\times$ 128                       \\ \cline{2-8} 
                                                                & 25                            & Transposed Conv                      & 64                                          & 2                                     & 2                                & ReLU                                          & 1280 $\times$ 64                       \\ \cline{2-8} 
                                                                & 26                            & \multicolumn{5}{c}{Addition (Layer 25, MHA with layer 3)}                                                                                                                                                     & 1280 $\times$ 64                       \\ \cline{2-8} 
                                                                & 27                            & TCN block with multiscale TIF module & 64                                          & Kernel: 3, Dilation rate: 16                     & 1                                & ReLU                                          & 1280 $\times$ 64                       \\ \cline{2-8} 
                                                                & 28                            & Transposed Conv                      & 64                                          & 2                                     & 2                                & ReLU                                          & 2560 $\times$ 64                       \\ \cline{2-8} 
                                                                & 29                            & Conv1D                               & 10                                          & 3                                     & 1                                & ReLU                                          & 2560 $\times$ 10                       \\ \hline
\end{tabular}}
\end{table*}

\blk
\blk
\subsection{Preprocessing}
ECG leads V1, V5, and lead  \uppercase\expandafter{\romannumeral2} are each represented by two records in the MITDB. Normally, lead  \uppercase\expandafter{\romannumeral2} is utilized in research to detect arrhythmia \cite{chen2020automated}. Since it is typically utilized for all patients, we only used lead \uppercase\expandafter{\romannumeral2} in the ECG signals. The ECG data were resampled at 256 Hz. After that, a Butterworth filter with three orders was used to adjust for baseline wandering. The ECG segment labels in the MITDB represent 2560 samples, each with 10-second recordings. Each sample is meant to have a class label.

No ECG segment values were used because each ECG data point has dissimilar amplitude scaling and vanishing offset effects. We employed normalization, which involved scaling the signals to the same level, to eliminate these effects. The Z score normalization method was used to normalize each ECG segment. The AFDB was subjected to an identical set of preprocessing steps a s that used for the MITDB.

\subsection{Step 1: Encoder}
\subsubsection{TCN Block with a Multiscale TIF Module as the Local-Global Temporal Component}

TCNs are substantially smaller than LSTM and have thus been constructed for sequence modeling tasks with a long effective history. As a result, we employed a TCN with a 10-s signal of size 2560 $\times$ 1 to capture local temporal dynamics by processing data in different time intervals. The TCN comprised two layers of dilated causal convolutions (dilated Conv1D), with layer normalization \cite{ioffe2015batch}, a rectified linear unit (ReLU) \cite{xu2015empirical}, and a drop-out layer \cite{srivastava2014dropout} in between the convolutions. An exponentially large receptive field was made possible by the dilated Conv1D \cite{yu2015multi}. Given a time sequence of input {$E$} $\in$ {R}$^n$ and filter {$f$} : $\lbrace$ 0, $\dots$, $k$ - 1 $\rbrace$ $\rightarrow{}$ {R}, the dilated Conv1D operation $T$ implemented on the element $t$ of a time step ${E}$ is defined as:

\begin{equation}
    T({E_{t}}) = ({E} \cdot d)(t) = \sum_{i=0}^{k-1} f(i) \cdot {E}_{t-d \cdot i}, {E} \leqq 0 := 0
\end{equation}

\begin{equation}
    Z = (T(E_{1}), T(E_{2}), \dots, T(E_{z}))
\end{equation}
where $d$ is the dilation rate, $k$ is the kernel size, $t$ - $d$ $\cdot$ $i$ counts the number of connections from previous nodes to the current node and $z$ is the output sequence with a length of $Z$.

The multiscale TIF module was composed of a series of three $10$ $\times$ $1$ convolution filters. In addition, the output of the second and third $10$ $\times$ $1$ convolution filters were concatenated and passed through $1$ $\times$ $1$ convolution filters. The sequence of large kernels with $10$ $\times$ $1$ filters enabled dense connections between the arrhythmic occurrence locations on the feature map for generating temporally stepwise semantic labels. Thus, we obtained multiscale temporal representations with three different temporal receptive fields, $i,e.,$ $10$ $\times$ $1$, $12$ $\times$ $1$, and $14$ $\times$ $1$. In addition, different large kernels could obtain useful information without incurring information losses by inspecting the points of interest from different scales. This approach enhanced the understanding of the arrhythmic global patterns. Therefore, the multiscale TIF module obtained global arrhythmia classification and localization information. Fig. 3 presents the architecture of the multiscale TIF module.

\begin{figure}
\begin{center}
\includegraphics[width = 0.9\columnwidth]{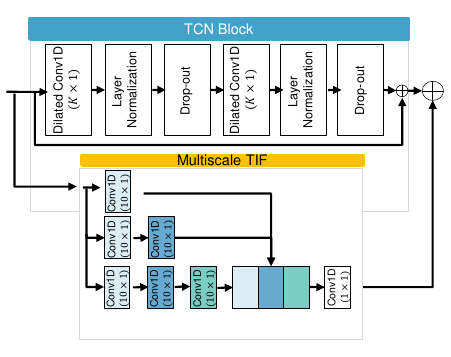}
\caption{Architecture of the TCN block with multiscale TIF module. The multiscale TIF consists of three Conv1Ds. Each Conv1D conducts $10$ $\times$ $1$ convolutions. One-dimensional dilated causal convolutional layer. Dilated Conv1D. Temporal information fusion. TIF.} 
\end{center}
\end{figure}

\blk 
\subsubsection{Temporal MHA Module}
The input sequences interacted with and were combined in a weighted manner for all time steps via the self-attention mechanism \cite{phyo2022transsleep}. The temporal MHA module was used to learn contextual representations of arrhythmia information by capturing temporal interactions. The MHA was used after applying the first-fifth TCN blocks with a multiscale TIF module. The MHA enabled the model to focus more on significant and pertinent time steps than on unimportant time steps from the sequential feature maps during recognition. Then, the output acquired from the MHA was added by utilizing the TCN with a multiscale TIF module in the decoder part. A mapping between a query $Q$ and sets of key-value pairs $K$ and $V$ can be used to express the attention function. The representation of the $i_{th}$ time step learned by the $h_{th}$ attention head is computed as:

\begin{equation}
    S = \frac{(W_{i}^Q) \cdot (W_{i}^K)^T} {\sqrt{{d_{k}}}})W_{i}^V
\end{equation}

\begin{equation}
    A = \frac{\exp (S)}{\sum_{y=1}^{i}\exp(S) }
\end{equation}

\begin{equation}
    h_{i} = \sum_{j=1}^{i}A\cdot W^V
\end{equation}

\begin{equation}
    M = (h_{i}, \dots, h_{h})W^0
\end{equation}

where $Q$, $K$, and $V$ are the matrices formed by the query, key, and value vectors, respectively, and $d$ is the scaling factor used to push the softmax function to locations where it has extremely small gradients. We used $Q$ = $K$ = $V$. The $W_{i}^Q$ $\in$ {R}$^{d  \times d_{k}}$, $W_{i}^K$ $\in$ {R}$^{d  \times d_{k}}$, and $W_{i}^V$ $\in$ {R}$^{d  \times d_{v}}$ are the projections as parameter matrices. $S$ can be expressed as scaled dot products. $A$ can be expressed as the scaled dot product attention. $h_{i}$ can be described as the computation of self-attention for each representation. $M$ can be expressed as the concatenation of the results of $h_{i}$ to $h_{h}$.

\subsection{Step 2: Bridge Part}
 Similar to the multiscale TIF module, this part used the comprehensive multiscale temporal semantics connecting the dilated Conv1D components in parallel to expand the receptive field series. This part was inspired by the bridge component in D-LinkNet \cite{zhou2018d}. To achieve multiscale feature fusion, the data from different branches were stacked in parallel using a variety of convolutional architectures with different dilation ratios. In the bridge part of the proposed method, dilated convolution layers with dilation rates of 1, 2, 4, and 8 were used. Therefore, the feature map used by the bridge part extracted the last layer in the encoder part. This part enabled the model to aggregate multiscale contextual information without losing it \cite{wu2019towards}. The bridge part covers 40 feature points of the feature map delivered through TCN with multiscale TIF. Therefore, this part all global information without a loss, including both the positional information at the start and end of the arrhythmia and the contextual information of the arrhythmia, and then transmits it to the decoder.

\subsection{Step 3: Decoder}

In this part, we constructed five TCNs with multiscale TIF modules and five transposed convolutions \cite{dumoulin2016guide}. The feature map $D_{in}$ after executing the bridge part, which was a transposed convolution layer, could be used to produce an upsampled output $D_{out}$ as follows:

\begin{equation}
\begin{split}
    D_{1} = D_{in} \circledast k_{1}\\
    D_{2} = D_{in} \circledast k_{2}\\
    D_{3} = D_{in} \circledast k_{3}\\
    D_{4} = D_{in} \circledast k_{4}\\
    D_{out} = D_{1} \oplus D_{2} \oplus D_{3} \oplus D_{4}
\end{split}
\end{equation}
where $\circledast$ denotes a convolutional operation, $\oplus$ denotes periodical shuffling, $k$ is a kernel, and $k_{i}$ for $i = 1, \dots, 4$. 

After the transposed convolution, upsampled outputs and feature map that passed through the temporal MHA were output to the TCN with the multiscale TIF module of the encoder and added. Then, the added feature map was passed through another TCN block with a multiscale TIF module to better understand the decoded contextual temporal information. The TCN with a multiscale TIF module operates in a similar manner as those in the encoder. The decoder first executed a transposed convolution, added the upsampled outputs, obtained a feature map from a temporal MHA mechanism, and finally passed the results through a TCN block with a multiscale TIF module. This sequence was repeated five times. Finally, the outputs were fed to a Conv1D layer after passing through a transposed convolution layer.

\subsection{Postprocessing}
We adopted an intelligible correction procedure to upgrade the raw arrhythmia predictions. In accordance with the descriptive statistics of the MITDB and AFDB, we knew that each arrhythmia event continued for a minimum of 256 samples. Therefore, when an event with a difference of less than 256 samples between its start and end points was found, we considered several options and proceeded with postprocessing according to each condition as follows.

First, when the arrhythmias before and after samples with a length below 256 were the same arrhythmia event, they were merged into a single event. For example, if AFIB were observed before and after a sequence of less than 256 samples, the result of this 10-s input was output as one AFIB event. On the other hand, if the events before and after an event containing less than 256 samples were different, the event with less than 256 samples was merged to the longer side among the occurrence times of the previous and subsequent events. For example, SBR occurred from 0 to 7 s and then changed to an AFIB event with less than 256 samples; in data where VT occurred from 7.8 s to 10 s, the AFIB event was corrected to SBR. 

\subsection{Model Training Details}

\subsubsection{Classwise Loss Function and Optimization}
We used and modified the Dice loss function \cite{sudre2017generalised} because it is the most common loss function for medical segmentation tasks \cite{sudre2017generalised}. Although the Dice loss function handles the imbalance problem, its ability to handle the imbalance in a classification task remains limited \cite{huang20203}. Moreover, we used CCE as another loss function because CCE individually evaluates the class predictions for each time step vector and then averages all samples. Therefore, we combined the modified Dice loss function with CCE to fit the multiclass scenario and assigned different weights to each class; thus, less frequent classes were assigned larger weights. This loss function is expressed as follows:

\begin{equation}
    CCE = - \sum_{{i=1}}^n y_{i,j} \log(p_{i,j})
\end{equation}

\begin{equation}
    L_{WDice} = 1 - \frac{2 \sum_{i=1}^n \sum_{j=1}^c w_{i,j}y_{i,j}p_{i,j}}{\sum_{i=1}^n \sum_{j=1}^c (y_{i,j} + p_{i,j}) \times w_{i,j}}
\end{equation}
where $i$ $\in$ $n$ is sample obtained in the $i$-th time step within the set of all time step samples $n$. $j$ $\in$ $c$ is the $j$-th class in the set of arrhythmias $c$. $p_{i,j}$ $\in$ $P$, $y_{i,j}$ $\in$ $Y$, and $w_{i,j}$ are the predicted probability, ground truth, and weight for class $j$ regarding the $i$-th time step sample.

The target weight $w_{i,j}$ is computed as follows:
\begin{equation}
    w_{i,j} = \frac{x^{tot}}{n \times x^{j}}
\end{equation}
where $x^{tot}$ is the total number of time step samples, and $n$ is the number of classes. $x^{j}$ is the number of time step samples in class $j$.

\begin{table}
\caption{Optimal Parameters of the Proposed Model for Detecting and Classifying Arrhythmias}
\centering
\renewcommand{\arraystretch}{1.2}
\begin{tabular}{cc}
\hline
\textbf{Hyperparameters}        & \textbf{Value} \\ \hline
Learning rate                   &   0.00005       \\ \hline
Optimizer                 &  Adam \cite{kingma2014adam}  \\ \hline
Batch size                      & 64  \\ \hline
\end{tabular}
\end{table}

\subsubsection{Model Interpretability}
To emphasize the regions in an arrhythmia prediction given a time step label \cite{selvaraju2017grad}, gradient-weighted class activation mapping (Grad-CAM) was used. This approach made the model predictions more understandable and enabled us to determine whether the labels assigned by the model were based on useful information \cite{degrave2021ai}. To implement Grad-CAM in the proposed model, we determined the gradients of the last stack of filters in each network for the relevant prediction class. These gradients allocated the significance of a specific time step for label prediction. Then, by conducting global average pooling on the gradients in each filter, we constructed filter importance weights by highlighting the filters whose gradient indicated that they contributed to the prediction of the class of interest. To create a Grad-CAM heatmap, each filter in the convolution layer of the final decoder was multiplied by its importance weight and aggregated across filters. The original ECG images were overlaid with the heatmaps.

\subsubsection{Implementational Details}
The batch size was set to 64, and the starting learning rate was set to 0.00005 with the learning rate decay technique. To update the weights, we utilized the adaptive moment estimation (Adam) optimizer \cite{kingma2014adam}. In the study, the convolution kernel size hyperparameters were modified. The above hyperparameters are listed in Table \uppercase\expandafter{\romannumeral2}.

The raw ECG data were loaded into MATLAB R2020b for preprocessing, and then Python 3.7 was utilized to implement the model with Keras v1.3.1. An NVIDIA RTX 8000 GPU was used for training and testing.

\section{Experiments and Validation}

\subsection{Database}

The experiments used two ECG datasets, which are summarized in Table \uppercase\expandafter{\romannumeral3}. The reasons for using these two ECG datasets were as follows. First, the MITDB and AFDB include multiple and multiclass arrhythmias. These datasets consist of data with durations exceeding 30 min or 10 h; thus, it was possible to make various arrhythmias exist or not exist in one input data sequence according to window segmentation. In contrast, most of the other ECG datasets consist of only presegmented input data and corresponding arrhythmia labels. In addition, these two datasets are most often used for arrhythmia classification studies, and they contain various arrhythmias, such as AFIB, AFL, VT, and supraventricular tachycardia (SVT). It was good to test the model performance in situations involving various arrhythmias. Although other long-term ECG databases are available, these datasets have only normal rhythms or binary classes, including AFIB and non-AFIB. Therefore, we tested the model performance in terms of solving one-input/multiple and multiclass output problems using only these two databases. Performances were evaluated by applying the proposed framework on each database, and the MITDB was utilized for the ablation study.

\begin{table}
\centering
\caption{Summary of Two Utilized Databases}
\renewcommand{\arraystretch}{1.2}
\resizebox{\columnwidth}{!}{
\begin{tabular}{ccccc}
\hline
\textbf{Database} & \textbf{\begin{tabular}[c]{@{}c@{}} No. of \\ Subject \end{tabular}} & \textbf{\begin{tabular}[c]{@{}c@{}} Sampling \\  Rate (Hz) \end{tabular}} & \textbf{\begin{tabular}[c]{@{}c@{}} Types of \\ Arrhythmia \end{tabular}} & \textbf{\begin{tabular}[c]{@{}c@{}} No. of \\ Samples \end{tabular}} \\ \hline
MITDB\cite{goldberger2000physiobank} & 47 & 360 & \begin{tabular}[c]{@{}c@{}} AFIB, AFL, \\ AVR, SVT, \\ VT, VF, PREX, \\ SBR, NSR, Other \end{tabular} & 13364 \\ \hline
AFDB\cite{goldberger2000physiobank} & 25 & 250 & \begin{tabular}[c]{@{}c@{}} AFIB, AFL, \\ AVR, NSR \end{tabular} & 74838 \\ \hline

\end{tabular}
}
\end{table}

\subsubsection{MITDB} Two-channel ambulatory ECG data were collected from 47 patients at the MIT-BIH Arrhythmia Laboratory to form this database. Each recording lasted for approximately 30 min. The sampling rate was 360 Hz. The recordings in the database correlate with 17 different types of arrhythmia and beats \cite{goldberger2000physiobank}. In this study, we chose to focus on 9 arrhythmic rhythm types out of the total of 17 since they had high levels of arrhythmia prevalence. The remaining classes were labeled as “other”, and 10 classes were categorized using the MITDB.

\subsubsection{AFDB} Two-channel ECG records for 25 patients with AFIB are available in the AFDB. Each data file lasts 10 h. A total of 250 samples per second were taken when sampling the signals. AFIB, AFL, AVR, and NSR are all included in the database.  \cite{goldberger2000physiobank}. Unlike the MITDB, we used all 4 classes, including AFIB, AFL, AVR, and NSR.


\subsection{$K$-fold Cross-Validation Setup}
For internal validation purposes, we employed $K$-fold cross-validation, which can address model overfitting issues. We utilized $K$ = 5 in this study because it is commonly used in many studies \cite{bugata2021feature}. The $5$-fold cross-validation method was used to shuffle the processed ECG segments at
random. The MITDB was employed for training with a rate of 80\%. The AFDB was organized similarly.

\subsection{Comparison Models} 
To validate (i) the performance of the proposed framework and (ii) its training and inference time, we compared the proposed model with UNet \cite{oh2019automated}, Res-UNet \cite{chen2022ru}, LinkNet \cite{chaurasia2017linknet}, and TCN-LinkNet (our baseline framework). We chose UNet \cite{oh2019automated}, Res-UNet \cite{chen2022ru} and LinkNet \cite{chaurasia2017linknet} for the comparison since these frameworks are widely used in the biomedical field. TCN-LinkNet indicates that the multiscale TIF, temporal MHA, and bridge parts of our proposed model were removed, and only TCNs were included in the encoder and decoder of LinkNet. We performed the same pre- and postprocessing steps for the same database according to the aforementioned loss function. 

To compare (iii) the generalization abilities of the models, we used the approach of Salinas et al. \cite{salinas2021detection}. Multiple and multiclass detection and classification for a given input are rarely addressed because many studies focus on multiclass arrhythmia classification with one input ECG dataset. Multiple detection and classification studies have mainly been conducted in binary classification cases, including those with AFIB and non-AFIB events. These studies used clinical cohort data; thus, we did not directly compare these studies. On the other hand, a one-input/multiple-binary task using a public dataset was reported only by Salinas et al. \cite{salinas2021detection}. To the best of our knowledge, we are the first to explore one-input/multiple and multiclass performance using raw ECG data. Therefore, we could not compare the one-input/multiple and multiclass performance of our approach due to the above reasons. Instead, to test and compare the generalization ability of our method through transfer learning, we tested and compared it using the MITDB, as in Salinas et al. \cite{salinas2021detection}.

\subsubsection{UNet \cite{oh2019automated}} UNet \cite{oh2019automated} consists of five compression stages that use a 1D convolutional layer and five upsampling states. The compressed feature maps in each compression stage concatenate the upsampled feature maps through skip connections. The skipped connections recover information that is lost during the compression stage. Therefore, UNet \cite{oh2019automated} is frequently used in the field of biosignals \cite{oh2019automated}.

\subsubsection{Res-UNet \cite{chen2022ru}} This method comprises five residual blocks as an encoder and four upsampling layers as a decoder. The encoder in Res-UNet \cite{chen2022ru} compresses features from a previous encoder block. The decoder in Res-UNet \cite{chen2022ru} consists of five upsampling layers that combine compressed feature maps and regular feature maps through skip connections. This method applies various time series segmentation approaches \cite{perslev2019u}.

\subsubsection{LinkNet \cite{chaurasia2017linknet}} Unlike Res-UNet \cite{chen2022ru}, LinkNet \cite{chaurasia2017linknet} combines features through addition instead of concatenation \cite{chaurasia2017linknet}. LinkNet \cite{chaurasia2017linknet} is an overall more efficient network because its decoder shares knowledge using fewer parameters. Therefore, this method is usable in the field of biosignals.

\subsubsection{TCN-LinkNet} TCN-LinkNet is the ablation framework used to test the multiscale TIF module in the TCN block, the bridge part, and the MHA in the input of each encoder layer of the proposed method. Therefore, TCN-LinkNet is composed of five TCN blocks as the encoder and five TCN blocks as the decoder.

\subsubsection{Salinas et al. \cite{salinas2021detection}} Salinas et al. \cite{salinas2021detection} proposed a CNN for the detection of AF (including AFIB and AFL) or non-AF (including rhythms without AF episodes), based on an electrocardio matrix. The architecture is composed of 3 convolutional layers, 3 batch normalization layers, 3 ReLU activation layers, 2 max pooling layers, 1 fully connected layer, 1 softmax layer, and 1 classification layer.

\subsection{Evaluation Metrics}
We used episode- and duration-based performance metrics \cite{sanders2016performance}. The performance metrics for the episode and duration tests were calculated using precision, recall, and the F1 score. The duration was calculated based on the temporal overlap between the period annotated by each classwise label and that detected and classified as each classwise label by the proposed method. We assessed the proportion of true episodes out of the total number of detected episodes that were approximately identified. We commonly computed the F1 score to compare duration and episode performance \cite{salinas2021detection}.

\subsubsection{Episode-Based Classification}

The episode-based performance evaluation criterion was the intersection over union (IoU). Each class label concerning the annotated period was compared to the overlap score between the detected and classified periods \cite{lea2017temporal}. The IoU score was classified as a true positive, false positive, or false negative if it fell below a threshold of 70\%. The number of properly predicted samples exceeding the IoU overlap criterion was $TP_{Epi_{i}}$, the number of erroneously predicted samples exceeding the IoU overlap criterion was $FP_{Epi_{i}}$, and the number of incorrectly predicted samples exceeding the IoU overlap criterion was $FN_{Epi_{i}}$. 

\subsubsection{Duration-Based Classification}
We calculated duration-based metrics by overlapping the annotated periods with the periods detected and classified by the proposed method. $TP_{Dur_{i}}$ is the total duration of the correctly predicted arrhythmia samples, $FP_{Dur_{i}}$ is the total duration of the incorrectly predicted arrhythmia samples, and $FN_{Epi_{i}}$ is the total duration of the incorrectly predicted nonarrhythmia samples.

\subsection{Statistical Analysis}

Kruskal‒Wallis tests were used to compare the performance metrics of the frameworks with and without each proposed component, and a post hoc analysis for assessing each component’s impact was performed using pairwise Wilcoxon tests with the least-significant difference (LSD) technique \cite{lee2022quantifying}. The Kruskal‒Wallis test and the Wilcoxon test with the LSD are widely used to determine which groups are significantly different \cite{marino2018statistical}. Additionally, the differences between various methods, UNet \cite{oh2019automated}, Res-UNet \cite{chen2022ru}, LinkNet \cite{chaurasia2017linknet}, the baseline of our proposed method and the proposed method, were evaluated using the Kruskal‒Wallis tests with the same method mentioned above for post hoc analyses. Next, the inference time differences among the proposed method and comparison methods, including UNet \cite{oh2019automated}, Res-UNet \cite{chen2022ru}, LinkNet \cite{chaurasia2017linknet}, and TCN-LinkNet, were assessed using Kruskal‒Wallis tests, and pairwise Wilcoxon tests with the LSD were applied for post hoc analyses. A significance level of 5\% ($p$ \textless{} 0.05) was considered for all analyses.

\section{Results and Discussion}

\subsection{Comparison Results on Different Databases}
The MITDB and AFDB were used to assess the performance of the proposed method. To solve the overfitting issue, we used $5$-fold cross-validation. Table \uppercase\expandafter{\romannumeral4} indicates the duration- and episode-based performances achieved by the proposed framework on the MITDB and AFDB. The 10-class duration- and episode-based performances achieved on the MITDB yielded overall F1 scores of 96.45\% and 82.05\%, respectively. The duration- and episode-based performances achieved on the AFDB in the 4-class setting produced F1 scores of 97.57\% and 98.31\%, respectively.

\begin{table}
\centering
\caption{Results of Duration- and Episode-Based Classification}
\renewcommand{\arraystretch}{1.2}
\begin{tabular}{cccc}
\hline
\multirow{2}{*}{\textbf{Database}} & \multirow{2}{*}{\textbf{Class}} & {\textbf{Duration}} & {\textbf{Episode}}                                               \\ \cline{3-4} 
                             & & \textbf{F1 score (\%)} &   \textbf{F1 score (\%)} \\  \hline
\multirow{10}{*}{\textbf{MITDB}} &  \textbf{AFIB}  &  94.11$\pm{}$2.09 & 77.64$\pm{}$1.78                      \\ \cline{2-4}
& \textbf{AFL}   &  92.91$\pm{}$2.12  & 83.64$\pm{}$1.51    \\  \cline{2-4}
& \textbf{AVR}   & 93.28$\pm{}$3.74 & 85.21$\pm{}$3.92 \\ \cline{2-4}
& \textbf{SVT}         & 91.76$\pm{}$6.43 & 72.85$\pm{}$6.84 \\ \cline{2-4}
& \textbf{SBR}   &  99.48$\pm{}$0.67   &  98.62$\pm{}$1.50  \\ \cline{2-4}
& \textbf{PREX}    & 93.67$\pm{}$1.28 & 61.22$\pm{}$3.39 \\ \cline{2-4}
& \textbf{VT}    & 92.54$\pm{}$3.35  & 84.50$\pm{}$3.33 \\ \cline{2-4}
& \textbf{VFIB}         & 93.99$\pm{}$2.92   & 79.20$\pm{}$3.43  \\ \cline{2-4}
& \textbf{NSR}     &  95.18$\pm{}$1.70  & 73.72$\pm{}$2.40  \\ \cline{2-4}
& \textbf{Other}      & 94.69$\pm{}$1.20   &  80.43$\pm{}$1.18 \\ \hline
\multirow{3}{*}{\textbf{AFDB}} & \textbf{AFIB}  & 98.45$\pm{}$0.72 & 98.39$\pm{}$0.74    \\ \cline{2-4}
& \textbf{AFL}  & 96.65$\pm{}$0.50  & 99.11$\pm{}$0.28 \\ \cline{2-4}
& \textbf{AVR}   & 96.01$\pm{}$2.91 & 97.10$\pm{}$1.92 \\ \cline{2-4}
& \textbf{NSR}   & 99.08$\pm{}$0.48 & 98.66$\pm{}$0.50  \\ \hline
\end{tabular}
\end{table}

\begin{table*}
\centering
\caption{Results of An Ablation Study Regarding the Effect of Each Component using the MITDB}
\renewcommand{\arraystretch}{1.2}
\begin{tabular}{ccccccccc}
\hline
\multirow{2}{*}{\textbf{Bridge block}} & \multirow{2}{*}{\textbf{ \textbf{\begin{tabular}[c]{@{}c@{}}MHA \\ without TIF \end{tabular}}}} &
\multirow{2}{*}{\textbf{\begin{tabular}[c]{@{}c@{}} Multiscale \\ TIF module \end{tabular}}} &\multicolumn{3}{c}{\textbf{Episode}} & \multicolumn{3}{c}{\textbf{Duration}} \\ \cline{4-9} 
                                             &         &                                 & \textbf{Precision (\%)}  & \textbf{Recall (\%)} & \textbf{F1 score (\%)} & \textbf{Precision (\%)}  & \textbf{Recall (\%)}  & \textbf{F1 score (\%)} \\ \hline
                                            &   &                                        &    69.64$\pm{2.53}$  &   87.74$\pm{3.61}$  &  76.95$\pm{2.78}$         &   94.24$\pm{0.84}$                  &      84.15$\pm{3.88}$            &         88.70$\pm{2.65}$           \\ \hline     
\checkmark                                            &  &                                         & 69.50$\pm{2.50}$  &  88.67$\pm{3.78}$  & 77.23$\pm{2.97}$      &  94.98$\pm{1.31}$                   & 85.34$\pm{3.92}$                 &     89.74$\pm{2.82}$                \\ \hline
                                            & \checkmark &                                         & 71.07$\pm{2.48}$ &   93.54$\pm{3.20}$ & 80.16$\pm{2.59}$   &   97.90$\pm{0.79}$                  & 91.78$\pm{3.66}$                 & 94.65$\pm{2.41}$   \\ \hline
                                            &      & \checkmark                                                     & 71.00$\pm{2.29}$  & 94.25$\pm{2.75}$ &    80.36$\pm{2.42}$               &   98.00$\pm{0.72}$                  &    92.80$\pm{3.11}$              &   95.28$\pm{2.03}$   \\ \hline

\checkmark                                            &  & \checkmark                                       & 72.87$\pm{2.78}$   &  94.41$\pm{2.64}$  &     81.58$\pm{2.45}$               &  98.54$\pm{0.40}$                   & 93.64$\pm{2.68}$                 & 95.96$\pm{1.64}$  \\ \hline

                                            & \checkmark & \checkmark                                       &  73.08$\pm{2.55}$   &  94.95$\pm{2.56}$ &     81.93$\pm{2.35}$              &   98.71$\pm{0.39}$                  &  94.23$\pm{2.63}$                &  96.37$\pm{1.60}$ \\ \hline

\checkmark                                            & \checkmark & \checkmark                                      & 73.18$\pm{2.53}$  & 95.12$\pm{2.68}$ &   82.05$\pm{2.38}$                & 98.79$\pm{0.31}$       & 94.33$\pm{2.82}$                 &  96.45$\pm{1.67}$   \\ \hline
\end{tabular}

{\scriptsize \checkmark \! indicates that the corresponding component was included. An empty space means that the corresponding component was excluded. Temporal information fusion: TIF. Multihead self-attention.}

\end{table*}

\subsection{Ablation Study}

\blk{The proposed framework is composed of three primary components: (i) a TCN block combined with a multiscale TIF module, (ii) an MHA mechanism derived from the output of the TCN block combined with a multiscale TIF module, and (iii) a bridge block using dilated Conv1D. We investigated the effects of these different components on the MITDB. To validate the different components, the following eight variant models were proposed. Table  \uppercase\expandafter{\romannumeral5} lists the overall duration- and episode-related precision, recall, and F1 score performances of the proposed components. 

Fig. 4 indicates that the multiscale TIF module significantly improved the overall duration and episode F1 scores. Compared with the baseline, although the MHA module exhibited no statistically significant improvements, the MHA module achieved overall score improvements of 5.95\%, and 3.21\%. The bridge partly slightly increased the duration- and episode-based performances.}

\subsubsection{Multiscale TIF for Extracting Global Pattern}
When we added the multiscale TIF module to the TCN block to fuse features derived from local temporal information and global patterns, the precision, recall, and F1 score values calculated for both the duration- and episode-based performances increased by more than 5\% over those of the TCN block alone, as shown in Table \uppercase\expandafter{\romannumeral5}. Furthermore, Fig. 4 indicates that the frameworks with the multiscale TIF module were statistically better than those without the multiscale TIF module. 

The reason for this finding is that the multiscale TIF module primarily captures different pieces of long-range global temporal information to fuse different resolutions of global patterns through multiscale large kernels \cite{szegedy2016rethinking, peng2017large}. As a result, the number of false negatives that were actually labeled but not detected by the algorithm decreased, and the recall metrics increased by approximately 6\% or more over those of the baseline in both the episode- and duration-based tests. In addition, the number of false positives that the algorithm erroneously detected was reduced, and the precision metric was also increased by approximately 1\%.

From the results and reasoning described above, we can determine that our proposed multiscale TIF module provides the proposed framework with advantages in terms of understanding and learning contextual arrhythmia information by extracting global patterns.

\begin{figure}
\begin{center}
\includegraphics[width = 1.0\columnwidth]{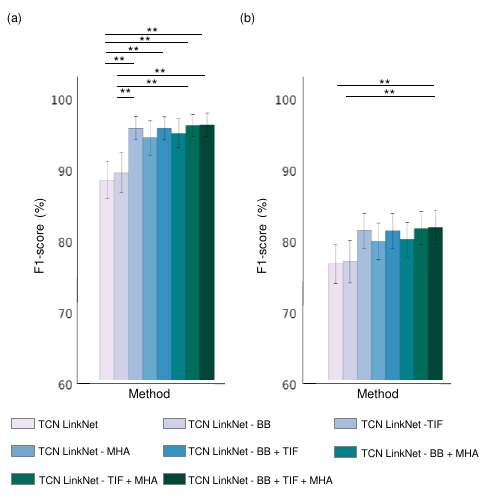}
\caption{Statistical comparison among the overall F1 scores produced by different components on the MITDB. (a) Overall duration F1 score. (b) Overall episode F1 score. The results are shown as overall means and $\pm{}$ standard deviations. ** $p$ \textless{} 0.01 (the Kruskal-Wallis test with the LSD was used for post hoc analyses).}
\end{center}
\end{figure}

\subsubsection{MHA Mechanism in the Output of the TCN Block without the Multiscale TIF Module}
We tested the effect of the MHA mechanism by removing the multiscale TIF module and bridge block (not the MHA). Table \uppercase\expandafter{\romannumeral5} shows that the MHA obtained precision, recall, and F1 score values of 71.07\%, 93.54\%, and 80.16\%, respectively, in the episode classification test. In addition, the precision, recall, and F1 score achieved in the duration classification test were 97.90\%, 91.78\%, and 94.65\%, respectively. Without the MHA or multiscale TIF module, our baseline framework yielded lower classification performances of 69.64\%, 87.74\%, and 76.95\% and 94.24\%, 84.15\%, and 88.70\%, respectively, for both the duration and episode classifications. The MHA module achieved higher values for all metrics than those of the baseline for both duration and episode classifications; however, the MHA module did not statistically differ from the baseline, as shown in Fig. 4. The MHA module delivers features that emphasize the distinctive information in the local temporal interaction features to the decoder, enabling the model to better understand and learn contextual information. Moreover, when the MHA module added the multiscale TIF module, all metrics were statistically enhanced compared with those of the baseline and the bridge component. Therefore, the MHA module satisfactorily captures distinct features within fused long-range local and global temporal information and transfers the corresponding TCN block with the multiscale TIF module in the decoder.

\subsubsection{Bridge Block using Dilated Conv1D}
Table \uppercase\expandafter{\romannumeral5} indicates that without the multiscale TIF module in the TCN block, the MHA in the output derived from the TCN block with TIF and the bridge block, TCN-LinkNet achieves precision, recall, and F1 score values of 69.64\%, 87.74\%, and 76.95\%, respectively, for episode-based classification. In addition, when the bridge block was added, the multiscale TIF module obtained precision, recall, and F1 score values of 69.50\%, 88.67\%, and 77.23\%, respectively, for episode-based classification. The bridge block yielded a performance improvement of less than 1\% over the baseline. However, when combined with the multiscale TIF module or the MHA, a performance improvement of more than 1\% was obtained. These results indicated that the bridge block plays a role in delivering the features extracted from the encoder of the framework to which each component is added with a different receptive field, and when feature fusion is performed, the contextual information of the features is delivered to the decoder without information loss \cite{wu2019towards}.

\subsubsection{Postprocessing}
We tested the effect of postprocessing using the MITDB. Supplementary Table \uppercase\expandafter{\romannumeral1} shows that the duration and episode classification performance achieved on the MITDB by the proposed method with postprocessing produced overall F1 scores of 96.45$\pm$1.67\% and 82.05$\pm$2.38\%, respectively. The overall F1 scores achieved for duration and episode classification by our proposed method without postprocessing were 96.60$\pm$1.63\% and 49.61$\pm$4.19\%, respectively. 

In summary, the effect of postprocessing was similar in terms of the overall duration F1 score, and the overall episode F1 score differed between the cases with and without postprocessing. The episode performance was calculated using counts of true positives, false negatives, and false positives \cite{sanders2016performance}. On the other hand, the duration performance was calculated based on the overlap between the ground-truth periods and the periods classified by our model. Regarding duration, there was no difference between the overall F1 scores produced with and without postprocessing because the overlapping periods between the ground truth and the proposed method were similar with and without postprocessing. However, the proposed method often output a fine region between arrhythmias, so this region was considered a false positive \cite{sanders2016performance}. These false positives worsened the episode performance of the model. As a result, postprocessing had no effect on duration performance but had an impact on episode performance. 

Finally, because the proposed method had an impact on episode performance, an advanced method for classifying fine areas is needed in the future.

\subsection{Comparison with Other Models}

\subsubsection{Performance}
We compared the performance of various models using 5-fold cross-validation on 10 classes of the MITDB and 4-class of the AFDB to investigate the effect of the proposed method. We used the same test database to conduct reasonable comparisons. In addition, we used the combination loss function of our proposed method for all the models.

The 10-class overall duration and episode F1 scores obtained by our proposed method on the MITDB were 94.65$\pm{}$1.67\% and 82.05$\pm{}$2.38\%, respectively. The overall F1 scores obtained by UNet \cite{oh2019automated} for the duration and episode tests were 78.22$\pm{}$2.19\% and 67.44$\pm{}$1.17\%, respectively. The overall F1 scores obtained by Res-UNet \cite{chen2022ru} for the duration and episode tests were 73.82$\pm{}$3.70\% and 63.84$\pm{}$2.83\%, respectively. LinkNet \cite{chaurasia2017linknet} achieved 69.21$\pm{}$3.31\% and 63.34$\pm{}$2.12\% F1 scores for the duration and episode tests, respectively. TCN-LinkNet (baseline) obtained 88.70$\pm{}$2.65\% and 76.95$\pm{}$2.78\% F1 scores for the duration and episode tests, respectively. Our proposed method obtained duration and episode F1 scores of 96.45$\pm{}$1.67\% and 82.05$\pm{}$2.38\%, respectively. The proposed method attained the best results in both the duration and episode performance tests when compared with the comparison models. Regarding the 4-class performance achieved on the AFDB, the proposed method also yielded the best results in both the duration and episode performance tests when compared with the other models, as shown in Table \uppercase\expandafter{\romannumeral6}.

Moreover, we performed the Kruskal‒Wallis test with the LSD post hoc test. Fig. 5 demonstrates the statistical analysis results produced by the comparison models, including UNet \cite{oh2019automated}, Res-UNet \cite{chen2022ru}, LinkNet \cite{chaurasia2017linknet}, TCN-LinkNet (baseline), and the proposed method. These results demonstrated that the proposed framework had statistically higher performance with respect to the episode and duration F1 scores ($p$ \textless{} 0.05).

The reasons for these higher performances are as follows. First, TCN-LinkNet performed local temporal information extraction \cite{lea2017temporal}. Therefore, TCN-LinkNet learned features that considered the interactions between the local time steps in the log range. As a result, compared with LinkNet, the overall duration and episode F1 scores improved by 19\% and 13\% on the MITDB. In addition, the overall duration F1 score improved by 10\% and the episode F1 score improved by 9\% when compared with those of UNet \cite{oh2019automated}. It is considered that the TCN-LinkNet extracting the contextual information between the local time step interactions overcomes the limitation of UNet \cite{oh2019automated}, which only uses local feature information through a convolution filter \cite{cai2022ma}. Second, our proposed method fused the local temporal interaction features and global pattern features by utilizing multiscale TIF to extract global pattern features into the skip connections of TCN-LinkNet. Furthermore, our proposed method added the MHA framework in TCN-LinkNet as our baseline model. Therefore, our proposed method could capture both local temporal information and different global temporal features, unlike UNet \cite{oh2019automated}, Res-UNet \cite{chen2022ru}, LinkNet \cite{chaurasia2017linknet}. As a result, duration and episode F1 scores increased by approximately 6\% over those of TCN-LinkNet.

\begin{table}
\centering
\caption{Comparison Among Overall F1 scores of the Models Using the MITDB and AFDB}
\renewcommand{\arraystretch}{1.2}
\resizebox{\columnwidth}{!}{
\begin{tabular}{ccccc}
\hline
\multirow{2}{*}{\textbf{Method}} & \multicolumn{2}{c}{\textbf{MITDB (10 class)}}               & \multicolumn{2}{c}{\textbf{AFDB (4 class)}}                \\ \cline{2-5} 
                                 & \textbf{ Duration (\%)} & \textbf{ Episode (\%)} & \textbf{Duration (\%)} & \textbf{Episode (\%)} \\ \hline
\textbf{UNet \cite{oh2019automated}}              &    78.22$\pm{2.19}$ &  67.44$\pm{1.17}$ &                     96.07$\pm{0.74}$   &  90.79$\pm{0.77}$    \\ \hline
\textbf{Res-UNet \cite{chen2022ru}}           & 73.82$\pm{3.70}$ & 63.84$\pm{2.83}$ & 91.24$\pm{1.20}$                       & 87.28$\pm{1.22}$ \\ \hline
\textbf{LinkNet \cite{chaurasia2017linknet}}                & 69.21$\pm{3.31}$ & 63.34$\pm{2.12}$   &      93.54$\pm{1.26}$                  & 89.11$\pm{0.67}$    \\ \hline
\textbf{TCN-LinkNet}                & 88.70$\pm{2.65}$ & 76.95$\pm{2.78}$   &   96.73$\pm{1.90}$    & 97.11$\pm{0.84}$ \\ \hline
\textbf{Proposed method}         &  \textbf{96.45}$\pm{\textbf{1.67}}$ & \textbf{82.05}$\pm{\textbf{2.38}}$  & \textbf{97.57}$\pm{\textbf{1.11}}$  & \textbf{98.31}$\pm{\textbf{0.76}}$ \\ \hline
\end{tabular}}
\end{table}

\begin{figure}[t]
\begin{center}
\includegraphics[width = 0.9\columnwidth]{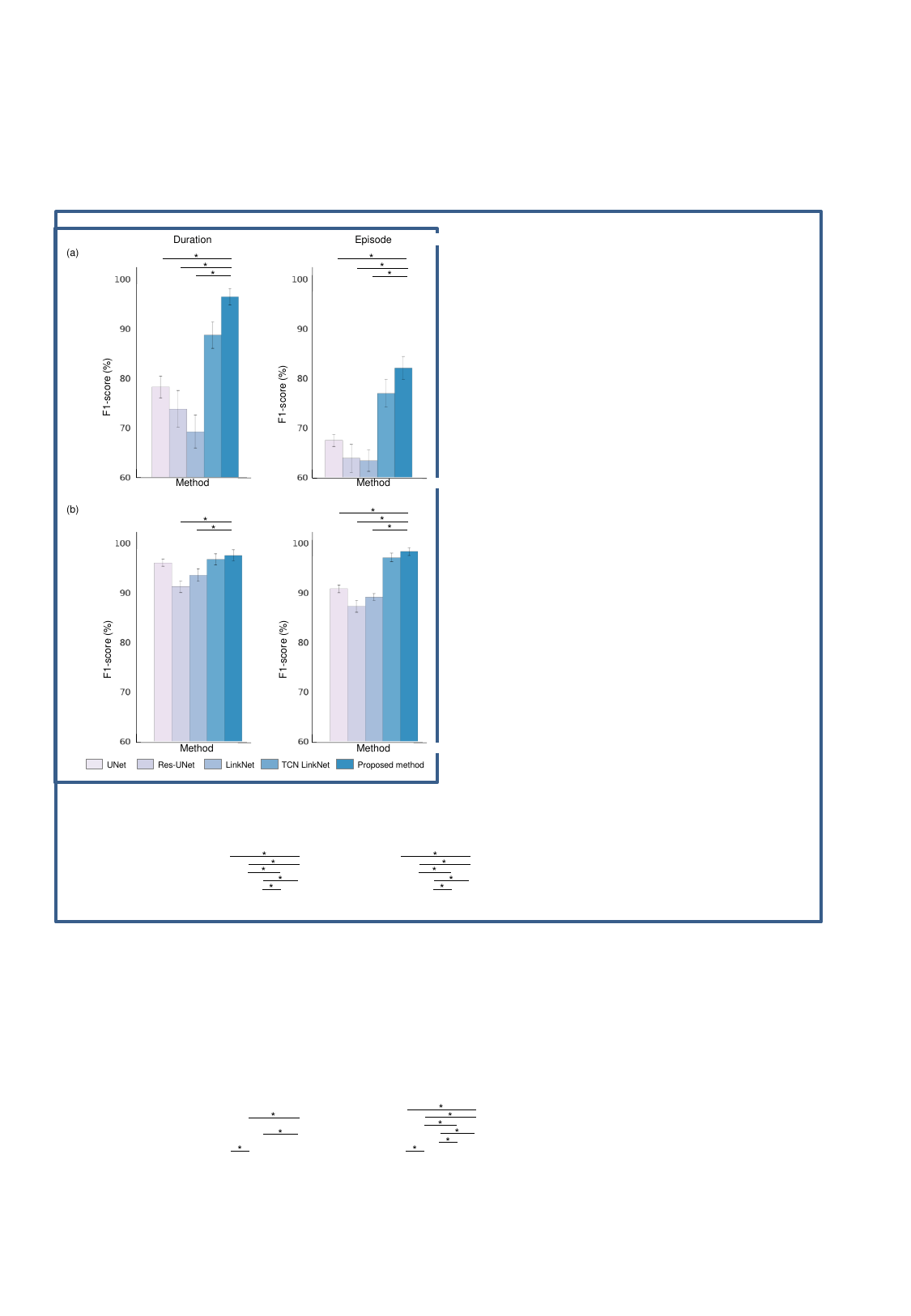}
\caption{Statistical comparison among the overall F1 score performances of the various models including UNet \cite{oh2019automated}, Res-UNet \cite{chen2022ru}, LinkNet \cite{chaurasia2017linknet}, TCN-LinkNet (our baseline), and the proposed method. (a) Overall duration and episode F1 scores achieved on the MITDB. (b) Overall duration and episode F1 scores achieved on the AFDB. We compared the models on two databases (the MITDB and AFDB). We only show the statistically significant differences between the proposed method and the comparison models. The results are presented as overall means $\pm{}$ standard deviations. * $p$ \textless{} 0.05 (the Kruskal‒Wallis test with the LSD was used for post hoc analyses). }
\end{center}
\end{figure}

\subsubsection{Training Time and Inference Time}
We analyzed the training times of the proposed model using the MITDB and compared them with those of the alternative models. Our proposed method achieved a training time per epoch of 87.50$\pm$0.55 s. The comparison models, UNet \cite{oh2019automated}, Res-UNet \cite{chen2022ru}, LinkNet \cite{chaurasia2017linknet}, TCN-LinkNet (Our baseline), obtained training times per epoch of 43.40$\pm$0.31 s, 58.30$\pm$0.30 s, 35.35$\pm$0.34 s, and 36.35$\pm$0.30 s, respectively. We analyzed the model inference times to check their applicability in the real world. The proposed model obtained a time of 0.50$\pm$0.01 s. The comparison models, UNet, Res-UNet, LinkNet, and TCN-LinkNet, obtained 0.11$\pm$0.01 s, 0.34$\pm$0.08 s, 0.11$\pm$0.01 s, and 0.22$\pm$0.01 s, respectively as shown in Supplementary Table \uppercase\expandafter{\romannumeral2} and  \uppercase\expandafter{\romannumeral3}. In summary, although our proposed method took longer to train per epoch than the alternatives because the proposed model is more complex than the other models, the inference time of the proposed method was not significantly different than those of the other methods. These findings suggest that the proposed method can be applied to inform clinicians of arrhythmia classification and detection information when the proposed model is used in CDSSs.

We analyzed the training times and inference times of the proposed model on the MITDB and compared them with those of the alternative methods. Although our proposed method took longer to train per epoch than the alternatives, the inference time of the proposed method was not significantly different than those of the other methods. Despite the fact that the proposed method can be applied to inform clinicians of clinical arrhythmia information in the real world, a more lightweight method is needed.

\subsubsection{Generalization Ability}

We tested the proposed framework on the AFDB as an independent database after training it on the MITDB as a source database to measure its generalization ability. The proposed method for classifying arrhythmias was compared to the approach of Salinas et al. \cite{salinas2021detection} in detail in Table \uppercase\expandafter{\romannumeral7}. We explain the comparison method and use the F1 score metric to evaluate the performance of each method.

Salinas et al. \cite{salinas2021detection} trained on a long-term atrial fibrillation database \cite{petrutiu2007abrupt} and the MIT-BIH NSR database \cite{goldberger2000physiobank} and then tested their method on the MITDB and AFDB. They obtained F1 scores of 45.52\% and 22.72\% regarding the duration and episode performances achieved on the MITDB, respectively. In addition, this method obtained F1 scores of 88.25\% and 75.11\% with respect to the duration and episode performances achieved on the AFDB. For a direct comparison, we trained on the AFDB, whose non-AF class included 10-s ECG signals from subjects with other cardiac arrhythmias as well as NSR subjects, whereas the AF class contained 10-s ECG signals from subjects with AFIB and AFL. Subsequently, we tested using the MITDB to calculate the F1 scores of the duration and episode performances. Consequently, we achieved F1 scores of 57.64\% and 64.44\% on the MITDB, respectively. Therefore, the results of multiple and multiclass AF and non-AF classifications performed using the proposed framework were higher than the results of Salinas et al. This indicated that the proposed method outperformed existing studies on the MITDB.

\begin{table}
\centering
\caption{Results of a Generalization Ability Comparison between the Proposed Model and Different Models}
\renewcommand{\arraystretch}{1.2}
\resizebox{\columnwidth}{!}{
\begin{tabular}{ccccccc}
\hline
\multirow{2}{*}{\textbf{Authors}} & \multirow{2}{*}{\textbf{Year}} & \multirow{2}{*}{\textbf{\begin{tabular}[c]{@{}c@{}}Training \\ Database\end{tabular}}} & \multirow{2}{*}{\textbf{\begin{tabular}[c]{@{}c@{}}Test \\ Database\end{tabular}}} & \multirow{2}{*}{\textbf{No. of Classes}} & \multicolumn{2}{c}{\textbf{F1 score (\%)}} \\ \cline{6-7} 
                                  &                                &                                                                                       &                                                                                   &                                        & \textbf{Duration}    & \textbf{Episode}    \\ \hline
Salinas et al. \cite{salinas2021detection}                    & 2021                           &  \begin{tabular}[c]{@{}c@{}} Long-term \\ atrial fibrillation \\ database, \\  MIT-BIH \\ normal sinus rhythm \\ database \end{tabular}                         & MITDB                                                                             & 2 & 45.52                & 22.72               \\ \hline
Proposed method                   & -                              & AFDB                                                                                  & MITDB                                                                             & 2       & \textbf{57.64}                & \textbf{64.44}               \\ \hline
\end{tabular}}

{\scriptsize MIT-BIH arrhythmia database, MITDB. MIT-BIH atrial fibrillation database, AFDB.}
\end{table}

Thus, the proposed framework successfully pulled features from multiple classes and learned even though the given input contained multiple arrhythmias. In real-world clinical environments, multiple arrhythmias can appear, and the classification of multiple arrhythmia within a single input is important. Furthermore, our model performed better than the state-of-the-art models in terms of arrhythmia duration and episode classification performance, indicating that it is more appropriate for real-world clinical environments using computational healthcare systems than the existing models.

\subsection{Clinical Interpretability}
Grad-CAM identified the arrhythmia sections that were most important for the time step label classification task. Fig. 6 highlights the capacity of our proposed method to extract temporal features that are related to clinical patterns. For NSR, the heatmap images indicated that our proposed method focused on P, R, and T waves. For arrhythmia, our proposed method focused on irregular QRS complexes. In particular, Grad-CAM highlighted an irregular complex, a T segment, and a fibrillation wave in the PREX and AFL cases. In addition, when multiple arrhythmias occurred, our proposed method stressed the importance of their onset and offset points between the current and next arrhythmias.  

Grad-CAM was used to highlight the prediction regions of multiple and multiclass labels. This provided the results predicted by the model with interpretability and enabled an evaluation of whether the predicted labels reflected clinically relevant information. This interpretability provides context for the predictions made by the proposed method and helps clinicians accept it in clinical workflows, as it can visually inform clinicians of the ECG parts used in AI-based ECG models that cannot be easily explained \cite{hughes2021performance}

\begin{figure}
\begin{center}
\includegraphics[width = 1.0\columnwidth]{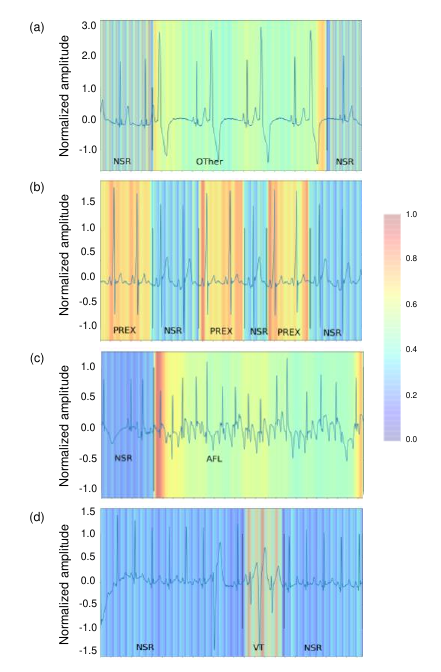}
\caption{\blk Heatmaps of the last convolution layer activations produced in the final decoder using Grad-CAM in the proposed method. (a) NSR and Other. (b) NSR and PREX. (c) NSR and AFL. (d) NSR and VT. The heatmaps of the activity in last convolution layer indicate that the proposed method distinguished multiple and multiclass arrhythmias. The color scale indicates the value of the time step. Our proposed method concentrated on aspects from blue to red and normalized them between 0 to 1. Red represents high arrhythmia significance, and blue signifies low arrhythmia significance for the proposed method.}
\end{center}
\end{figure}

\section{Conclusion}
In this study, we presented a novel CDSS that classifies multiple and multiclass arrhythmia based on the proposed local-global temporal fusion network with an attention mechanism. The proposed framework includes a TCN with a multiscale TIF module as an encoder and decoder, a temporal MHA module, and a bridge part comprising a dilated convolutional block using one-lead ECG data. The proposed framework enhances the ability to detect and classify multiple classes through two novel strategies: combining the TCN with multiscale MHA and combining the temporal MHA in the encoder with the upsampled features in the decoder. Moreover, the proposed framework uses time shifting as an augmentation method and a combination of the multiclass weighted Dice loss function and CCE to resolve the imbalance problem among class categories. Therefore, the proposed framework could localize and classify the multiple and multiclass arrhythmia in the given input. Consequently, it is possible to exploit multiple and multiclass classification models that exhibit stable classification performance to calculate the duration of arrhythmia and classify it for practical use in the future.  

The results showed that the proposed model outperformed UNet \cite{oh2019automated}, Res-UNet \cite{chen2022ru}, LinkNet \cite{chaurasia2017linknet}, and TCN-LinkNet (the baseline framework of the proposed method) in terms of all evaluation metrics. In addition, an ablation study was performed to evaluate the effect of each component of the architecture. The multiscale TIF framework exhibited a statistically significant improvement over TCN-LinkNet and the TCN-LinkNet version with the bridge part. Although the temporal MHA framework did not yield a statistically significant improvement over TCN-LinkNet, improvements of 3\% and 5\% was observed for the episode and duration performances, respectively. The bridge part slightly improved the results when combined with the multiscale TIF and temporal MHA modules. The combination of the multiscale TIF module, the temporal MHA module, and the bridge part outperformed the state-of-the-art models in multiple arrhythmia detection and classification evaluations concerning the duration and episode F1 scores achieved on the MITDB. It can help clinicians incorporate predictive information into clinical workflows by providing visual representations of the information used by AI-based ECG models. Consequently, we confirmed that the TCN-LinkNet model combined with the multiscale TIF module, the temporal MHA module, and the bridge part architecture detected and classified multiple and multiclass arrhythmia while capturing local and global arrhythmia patterns in one-lead ECG signals. Finally, the proposed framework can be useful for calculating episodes and durations in a CDSS. Cardiologists may use it to plan diagnoses and treatments accordingly in clinical settings. 

\bibliography{Reference.bib}{}

\begin{thebibliography}{10}
\providecommand{\url}[1]{#1}
\csname url@samestyle\endcsname
\providecommand{\newblock}{\relax}
\providecommand{\bibinfo}[2]{#2}
\providecommand{\BIBentrySTDinterwordspacing}{\spaceskip=0pt\relax}
\providecommand{\BIBentryALTinterwordstretchfactor}{4}
\providecommand{\BIBentryALTinterwordspacing}{\spaceskip=\fontdimen2\font plus
\BIBentryALTinterwordstretchfactor\fontdimen3\font minus
  \fontdimen4\font\relax}
\providecommand{\BIBforeignlanguage}[2]{{%
\expandafter\ifx\csname l@#1\endcsname\relax
\typeout{** WARNING: IEEEtran.bst: No hyphenation pattern has been}%
\typeout{** loaded for the language `#1'. Using the pattern for}%
\typeout{** the default language instead.}%
\else
\language=\csname l@#1\endcsname
\fi
#2}}
\providecommand{\BIBdecl}{\relax}
\BIBdecl

\bibitem{xu2021comorbidity}
Z.~Xu, J.~Zhang, Q.~Zhang, Q.~Xuan, and P.~S.~F. Yip, ``A comorbidity
  knowledge-aware model for disease prognostic prediction,'' \emph{IEEE Trans.
  Cybern.}, 2021.

\bibitem{jeong2022real}
J.-H. Jeong, J.-H. Cho, B.-H. Lee, and S.-W. Lee, ``Real-time deep
  neurolinguistic learning enhances noninvasive neural language decoding for
  brain-machine interaction,'' \emph{IEEE Trans. Cybern.}, 2022.

\bibitem{sutton2020overview}
R.~T. Sutton, D.~Pincock, D.~C. Baumgart, D.~C. Sadowski, R.~N. Fedorak, and
  K.~I. Kroeker, ``An overview of clinical decision support systems: benefits,
  risks, and strategies for success,'' \emph{NPJ Digit.}, vol.~3, no.~1, pp.
  1--10, 2020.

\bibitem{sidek2014ecg}
K.~A. Sidek, I.~Khalil, and H.~F. Jelinek, ``{ECG} biometric with abnormal
  cardiac conditions in remote monitoring system,'' \emph{IEEE Trans. Syst. Man
  Cybern. Syst.}, vol.~44, no.~11, pp. 1498--1509, 2014.

\bibitem{kamel2013paroxysmal}
H.~Kamel, M.~S. Elkind, P.~D. Bhave, B.~B. Navi, P.~M. Okin, C.~Iadecola, R.~B.
  Devereux, and M.~E. Fink, ``Paroxysmal supraventricular tachycardia and the
  risk of ischemic stroke,'' \emph{Stroke}, vol.~44, no.~6, pp. 1550--1554,
  2013.

\bibitem{atlee2006complications}
J.~L. Atlee, \emph{Complications in Anesthesia E-Book}.\hskip 1em plus 0.5em
  minus 0.4em\relax Elsevier Health Sciences, 2006.

\bibitem{sannino2018deep}
G.~Sannino and G.~De~Pietro, ``A deep learning approach for {ECG}-based
  heartbeat classification for arrhythmia detection,'' \emph{Future Gener.
  Comp. Syst.}, vol.~86, pp. 446--455, 2018.

\bibitem{kim2022automatic}
Y.~K. Kim, M.~Lee, H.~S. Song, and S.-W. Lee, ``Automatic cardiac arrhythmia
  classification using residual network combined with long short-term memory,''
  \emph{IEEE Trans. Instrum. Meas.}, 2022.

\bibitem{huang2014new}
H.~Huang, J.~Liu, Q.~Zhu, R.~Wang, and G.~Hu, ``A new hierarchical method for
  inter-patient heartbeat classification using random projections and {RR}
  intervals,'' \emph{Biomed. Eng. Online}, vol.~13, no.~1, pp. 1--26, 2014.

\bibitem{pourbabaee2018deep}
B.~Pourbabaee, M.~J. Roshtkhari, and K.~Khorasani, ``Deep convolutional neural
  networks and learning {ECG} features for screening paroxysmal atrial
  fibrillation patients,'' \emph{IEEE Trans. Syst. Man Cybern. Syst.}, vol.~48,
  no.~12, pp. 2095--2104, 2018.

\bibitem{acharya2017automated}
U.~R. Acharya, H.~Fujita, O.~S. Lih, Y.~Hagiwara, J.~H. Tan, and M.~Adam,
  ``Automated detection of arrhythmias using different intervals of tachycardia
  {ECG} segments with convolutional neural network,'' \emph{Inf. Sci.}, vol.
  405, pp. 81--90, 2017.

\bibitem{faust2018automated}
O.~Faust, A.~Shenfield, M.~Kareem, T.~R. San, H.~Fujita, and U.~R. Acharya,
  ``Automated detection of atrial fibrillation using long short-term memory
  network with {RR} interval signals,'' \emph{Comput. Biol. Med.}, vol. 102,
  pp. 327--335, 2018.

\bibitem{chen2020automated}
C.~Chen, Z.~Hua, R.~Zhang, G.~Liu, and W.~Wen, ``Automated arrhythmia
  classification based on a combination network of {CNN} and {LSTM},''
  \emph{Biomed. Signal. Process.}, vol.~57, p. 101819, 2020.

\bibitem{teplitzky2020deep}
B.~A. Teplitzky, M.~McRoberts, and H.~Ghanbari, ``Deep learning for
  comprehensive {ECG} annotation,'' \emph{Heart rhythm}, vol.~17, no.~5, pp.
  881--888, 2020.

\bibitem{zhang2019online}
L.~Zhang, J.~Zhao, and W.~Li, ``Online and unsupervised anomaly detection for
  streaming data using an array of sliding windows and pdds,'' \emph{IEEE
  Trans. Cybern.}, vol.~51, no.~4, pp. 2284--2289, 2019.

\bibitem{aly2005survey}
M.~Aly, ``Survey on multiclass classification methods,'' \emph{Neural Netw.},
  vol.~19, no. 1-9, p.~2, 2005.

\bibitem{parvaneh2019cardiac}
S.~Parvaneh, J.~Rubin, S.~Babaeizadeh, and M.~Xu-Wilson, ``Cardiac arrhythmia
  detection using deep learning: A review,'' \emph{J. Electrocardiol.},
  vol.~57, pp. S70--S74, 2019.

\bibitem{miyasaka2006secular}
Miyasaka, ``Secular trends in incidence of atrial fibrillation in olmsted
  county, minnesota, 1980 to 2000, and implications on the projections for
  future prevalence,'' \emph{Circulation}, vol. 114, no.~11, pp. E498--E498,
  2006.

\bibitem{khurshid2018frequency}
S.~Khurshid, S.~H. Choi, L.-C. Weng, E.~Y. Wang, L.~Trinquart, E.~J. Benjamin,
  P.~T. Ellinor, and S.~A. Lubitz, ``Frequency of cardiac rhythm abnormalities
  in a half million adults,'' \emph{Circ. Arrhythm. Electrophysiol.}, vol.~11,
  no.~7, p. e006273, 2018.

\bibitem{walsh2007arrhythmias}
E.~P. Walsh and F.~Cecchin, ``Arrhythmias in adult patients with congenital
  heart disease,'' \emph{Circulation}, vol. 115, no.~4, pp. 534--545, 2007.

\bibitem{chaurasia2017linknet}
A.~Chaurasia and E.~Culurciello, ``Linknet: Exploiting encoder representations
  for efficient semantic segmentation,'' in \emph{IEEE Visual Communications
  and Image Processing}.\hskip 1em plus 0.5em minus 0.4em\relax IEEE, 2017, pp.
  1--4.

\bibitem{lea2017temporal}
C.~Lea, M.~D. Flynn, R.~Vidal, A.~Reiter, and G.~D. Hager, ``Temporal
  convolutional networks for action segmentation and detection,'' in \emph{IEEE
  Conference on Computer Vision and Pattern Recognition}, 2017, pp. 156--165.

\bibitem{yu2015multi}
F.~Yu and V.~Koltun, ``Multi-scale context aggregation by dilated
  convolutions,'' \emph{arXiv preprint arXiv:1511.07122}, 2015.

\bibitem{vaswani2017attention}
A.~Vaswani, N.~Shazeer, N.~Parmar, J.~Uszkoreit, L.~Jones, A.~N. Gomez,
  {\L}.~Kaiser, and I.~Polosukhin, ``Attention is all you need,'' \emph{Adv.
  Neural Inf. Process. Syst.}, vol.~30, 2017.

\bibitem{moody2001impact}
G.~B. Moody and R.~G. Mark, ``The impact of the {MIT}-{BIH} arrhythmia
  database,'' \emph{IEEE Eng. Med. Biol.}, vol.~20, no.~3, pp. 45--50, 2001.

\bibitem{khalifa2019character}
M.~Khalifa and K.~Shaalan, ``Character convolutions for arabic named entity
  recognition with long short-term memory networks,'' \emph{Comput. Speech
  Lang.}, vol.~58, pp. 335--346, 2019.

\bibitem{he2019automatic}
R.~He, Y.~Liu, K.~Wang, N.~Zhao, Y.~Yuan, Q.~Li, and H.~Zhang, ``Automatic
  cardiac arrhythmia classification using combination of deep residual network
  and bidirectional {LSTM},'' \emph{IEEE Access}, vol.~7, pp.
  102\,119--102\,135, 2019.

\bibitem{petryshak2021robust}
B.~Petryshak, I.~Kachko, M.~Maksymenko, and O.~Dobosevych, ``Robust deep
  learning pipeline for {PVC} beats localization,'' \emph{Technol. Health
  Care}, vol.~29, no.~S1, pp. 475--486, 2021.

\bibitem{johnson2017google}
M.~Johnson, M.~Schuster, Q.~V. Le, M.~Krikun, Y.~Wu, Z.~Chen, N.~Thorat,
  F.~Vi{\'e}gas, M.~Wattenberg, G.~Corrado \emph{et~al.}, ``Google’s
  multilingual neural machine translation system: Enabling zero-shot
  translation,'' \emph{Trans. Assoc. Comput.}, vol.~5, pp. 339--351, 2017.

\bibitem{pokaprakarn2021sequence}
T.~Pokaprakarn, R.~R. Kitzmiller, R.~Moorman, D.~E. Lake, A.~K. Krishnamurthy,
  and M.~R. Kosorok, ``Sequence to sequence ecg cardiac rhythm classification
  using convolutional recurrent neural networks,'' \emph{IEEE J. Biomed. Health
  Inform.}, vol.~26, no.~2, pp. 572--580, 2021.

\bibitem{bai2018empirical}
S.~Bai, J.~Z. Kolter, and V.~Koltun, ``An empirical evaluation of generic
  convolutional and recurrent networks for sequence modeling,'' \emph{arXiv
  preprint arXiv:1803.01271}, 2018.

\bibitem{ma2021ecg}
H.~Ma, C.~Chen, Q.~Zhu, H.~Yuan, L.~Chen, and M.~Shu, ``An {ECG} signal
  classification method based on dilated causal convolution,'' \emph{Comput.
  Math. Methods Med.}, vol. 2021, 2021.

\bibitem{ingolfsson2021ecg}
T.~M. Ingolfsson, X.~Wang, M.~Hersche, A.~Burrello, L.~Cavigelli, and
  L.~Benini, ``{ECG}-{TCN}: Wearable cardiac arrhythmia detection with a
  temporal convolutional network,'' in \emph{Artificial Intelligence Circuits
  and Systems}.\hskip 1em plus 0.5em minus 0.4em\relax IEEE, 2021, pp. 1--4.

\bibitem{ibtehaz2020multiresunet}
N.~Ibtehaz and M.~S. Rahman, ``Multiresunet: Rethinking the {U-Net}
  architecture for multimodal biomedical image segmentation,'' \emph{Neural
  Netw.}, vol. 121, pp. 74--87, 2020.

\bibitem{peng2017large}
C.~Peng, X.~Zhang, G.~Yu, G.~Luo, and J.~Sun, ``Large kernel matters--improve
  semantic segmentation by global convolutional network,'' in \emph{IEEE
  Conference on Computer Vision and Pattern Recognition}, 2017, pp. 4353--4361.

\bibitem{ioffe2015batch}
S.~Ioffe and C.~Szegedy, ``Batch normalization: Accelerating deep network
  training by reducing internal covariate shift,'' in \emph{International
  Conference on Machine Learning}.\hskip 1em plus 0.5em minus 0.4em\relax PMLR,
  2015, pp. 448--456.

\bibitem{xu2015empirical}
B.~Xu, N.~Wang, T.~Chen, and M.~Li, ``Empirical evaluation of rectified
  activations in convolutional network,'' \emph{arXiv preprint
  arXiv:1505.00853}, 2015.

\bibitem{srivastava2014dropout}
N.~Srivastava, G.~Hinton, A.~Krizhevsky, I.~Sutskever, and R.~Salakhutdinov,
  ``Dropout: a simple way to prevent neural networks from overfitting,''
  \emph{J. Mach. Learn. Res.}, vol.~15, no.~1, pp. 1929--1958, 2014.

\bibitem{phyo2022transsleep}
J.~Phyo, W.~Ko, E.~Jeon, and H.-I. Suk, ``Transsleep: Transitioning-aware
  attention-based deep neural network for sleep staging,'' \emph{IEEE Trans.
  Cybern.}, 2022.

\bibitem{zhou2018d}
L.~Zhou, C.~Zhang, and M.~Wu, ``D-{L}ink{N}et: Link{N}et with pretrained
  encoder and dilated convolution for high resolution satellite imagery road
  extraction,'' in \emph{IEEE Conference on Computer Vision and Pattern
  Recognition}, 2018, pp. 182--186.

\bibitem{wu2019towards}
M.~Wu, C.~Zhang, J.~Liu, L.~Zhou, and X.~Li, ``Towards accurate high resolution
  satellite image semantic segmentation,'' \emph{IEEE Access}, vol.~7, pp.
  55\,609--55\,619, 2019.

\bibitem{dumoulin2016guide}
V.~Dumoulin and F.~Visin, ``A guide to convolution arithmetic for deep
  learning,'' \emph{arXiv preprint arXiv:1603.07285}, 2016.

\bibitem{sudre2017generalised}
C.~H. Sudre, W.~Li, T.~Vercauteren, S.~Ourselin, and M.~Jorge~Cardoso,
  ``Generalised dice overlap as a deep learning loss function for highly
  unbalanced segmentations,'' in \emph{Deep Learning in Medical Image Analysis
  and Multimodal Learning for Clinical Decision Support}.\hskip 1em plus 0.5em
  minus 0.4em\relax Springer, 2017, pp. 240--248.

\bibitem{huang20203}
Y.-J. Huang, Q.~Dou, Z.-X. Wang, L.-Z. Liu, Y.~Jin, C.-F. Li, L.~Wang, H.~Chen,
  and R.-H. Xu, ``3-{D} {R}o{I}-aware {U-N}et for accurate and efficient
  colorectal tumor segmentation,'' \emph{IEEE Trans. Cybern.}, vol.~51, no.~11,
  pp. 5397--5408, 2020.

\bibitem{kingma2014adam}
D.~P. Kingma and J.~Ba, ``Adam: A method for stochastic optimization,''
  \emph{arXiv preprint arXiv:1412.6980}, 2014.

\bibitem{selvaraju2017grad}
R.~R. Selvaraju, M.~Cogswell, A.~Das, R.~Vedantam, D.~Parikh, and D.~Batra,
  ``Grad-{CAM}: Visual explanations from deep networks via gradient-based
  localization,'' in \emph{International Conference on Computer Vision}, 2017,
  pp. 618--626.

\bibitem{degrave2021ai}
A.~J. DeGrave, J.~D. Janizek, and S.-I. Lee, ``{AI} for radiographic {COVID}-19
  detection selects shortcuts over signal,'' \emph{Nat. Mach. Intell.}, vol.~3,
  no.~7, pp. 610--619, 2021.

\bibitem{goldberger2000physiobank}
A.~L. Goldberger, L.~A. Amaral, L.~Glass, J.~M. Hausdorff, P.~C. Ivanov, R.~G.
  Mark, J.~E. Mietus, G.~B. Moody, C.-K. Peng, and H.~E. Stanley, ``Physiobank,
  physiotoolkit, and physionet: components of a new research resource for
  complex physiologic signals,'' \emph{Circulation}, vol. 101, no.~23, pp.
  e215--e220, 2000.

\bibitem{bugata2021feature}
P.~Bugata and P.~Drotar, ``Feature selection based on a sparse neural-network
  layer with normalizing constraints,'' \emph{IEEE Trans. Cybern.}, 2021.

\bibitem{oh2019automated}
S.~L. Oh, E.~Y. Ng, R.~San~Tan, and U.~R. Acharya, ``Automated beat-wise
  arrhythmia diagnosis using modified {U-N}et on extended electrocardiographic
  recordings with heterogeneous arrhythmia types,'' \emph{Comput. Biol. Med.},
  vol. 105, pp. 92--101, 2019.

\bibitem{chen2022ru}
J.~Chen, H.~Deng, S.~Li, W.~Li, H.~Chen, Y.~Chen, and B.~Luo, ``{RU}-{N}et: A
  residual {U-Net} for automatic interplanetary coronal mass ejection
  detection,'' \emph{Astrophys. J., Suppl. Ser.}, vol. 259, no.~1, p.~8, 2022.

\bibitem{salinas2021detection}
R.~Salinas-Mart{\'\i}nez, J.~De~Bie, N.~Marzocchi, and F.~Sandberg, ``Detection
  of brief episodes of atrial fibrillation based on electrocardiomatrix and
  convolutional neural network,'' \emph{Front. Physiol.}, vol.~12, 2021.

\bibitem{perslev2019u}
M.~Perslev, M.~Jensen, S.~Darkner, P.~J. Jennum, and C.~Igel, ``U-time: A fully
  convolutional network for time series segmentation applied to sleep
  staging,'' \emph{Adv. Neural Inf. Process. Syst.}, vol.~32, 2019.

\bibitem{sanders2016performance}
P.~Sanders, H.~P{\"u}rerfellner, E.~Pokushalov, S.~Sarkar, M.~Di~Bacco,
  B.~Maus, L.~R. Dekker, R.~L.~U. Investigators \emph{et~al.}, ``Performance of
  a new atrial fibrillation detection algorithm in a miniaturized insertable
  cardiac monitor: Results from the reveal {LINQ} usability study,''
  \emph{Heart Rhythm}, vol.~13, no.~7, pp. 1425--1430, 2016.

\bibitem{lee2022quantifying}
M.~Lee, L.~R. Sanz, A.~Barra, A.~Wolff, J.~O. Nieminen, M.~Boly, M.~Rosanova,
  S.~Casarotto, O.~Bodart, J.~Annen \emph{et~al.}, ``Quantifying arousal and
  awareness in altered states of consciousness using interpretable deep
  learning,'' \emph{Nat. Commun.}, vol.~13, no.~1, p. 1064, 2022.

\bibitem{marino2018statistical}
M.~J. Marino, ``Statistical analysis in preclinical biomedical research,'' in
  \emph{Research in the Biomedical Sciences}.\hskip 1em plus 0.5em minus
  0.4em\relax Elsevier, 2018, pp. 107--144.

\bibitem{szegedy2016rethinking}
C.~Szegedy, V.~Vanhoucke, S.~Ioffe, J.~Shlens, and Z.~Wojna, ``Rethinking the
  inception architecture for computer vision,'' in \emph{IEEE Conference on
  Computer Vision and Pattern Recognition}, 2016, pp. 2818--2826.

\bibitem{cai2022ma}
Y.~Cai and Y.~Wang, ``{MA-UN}et: An improved version of {UN}et based on
  multi-scale and attention mechanism for medical image segmentation,'' in
  \emph{International Conference on Electronics and Communication; Network and
  Computer Technology}, vol. 12167.\hskip 1em plus 0.5em minus 0.4em\relax
  SPIE, 2022, pp. 205--211.

\bibitem{petrutiu2007abrupt}
S.~Petrutiu, A.~V. Sahakian, and S.~Swiryn, ``Abrupt changes in fibrillatory
  wave characteristics at the termination of paroxysmal atrial fibrillation in
  humans,'' \emph{Europace}, vol.~9, no.~7, pp. 466--470, 2007.

\bibitem{hughes2021performance}
J.~W. Hughes, J.~E. Olgin, R.~Avram, S.~A. Abreau, T.~Sittler, K.~Radia,
  H.~Hsia, T.~Walters, B.~Lee, J.~E. Gonzalez \emph{et~al.}, ``Performance of a
  convolutional neural network and explainability technique for 12-lead
  electrocardiogram interpretation,'' \emph{JAMA Cardiol.}, vol.~6, no.~11, pp.
  1285--1295, 2021.

\end{thebibliography}
\bibliographystyle{IEEEtran}

\end{document}